\documentclass{article}
\usepackage[utf8]{inputenc}
\usepackage{graphicx}
\usepackage{geometry}
\usepackage{authblk}
\usepackage{amsmath}
\usepackage{amssymb}
\usepackage{xcolor}
\usepackage{graphbox}
\usepackage{hyperref}
\newcommand{\N}{\ensuremath{\mathbb{N}}}
\usepackage[linesnumbered,ruled]{algorithm2e}
\usepackage{enumitem}
\usepackage{xcolor,soul}
\usepackage[sorting = none, backend = bibtex, style=numeric-comp]{biblatex}

\title{Noise Dynamics of Quantum Annealers: Estimating the Effective Noise Using Idle Qubits}

\author[1]{Elijah Pelofske \footnote{epelofske@lanl.gov}}
\author[2]{Georg Hahn}
\author[1,3]{Hristo N.\ Djidjev}

\affil[1]{Los Alamos National Laboratory, CCS-3, Los Alamos, NM 87545, USA}
\affil[2]{Harvard T.H.\ Chan School of Public Health, Boston, MA 02115, USA}
\affil[3]{Institute of Information and Communication Technologies, Bulgarian Academy of Sciences, Sofia, Bulgaria}

\date{\vspace{-6ex}}

\begin{document}
\maketitle

\begin{abstract}
Quantum annealing is a type of analog computation that aims to use quantum mechanical fluctuations in search of optimal solutions of QUBO (quadratic unconstrained binary optimization) or, equivalently, Ising problems. Since NP-hard problems can in general be mapped to Ising and QUBO formulations, the quantum annealing paradigm has the potential to help solve various NP-hard problems. Current quantum annealers, such as those manufactured by D-Wave Systems, Inc., have various practical limitations including the size (number of qubits) of the problem that can be solved, the qubit connectivity, and error due to the environment or system calibration, which can reduce the quality of the solutions. Typically, for an arbitrary problem instance, the corresponding QUBO (or Ising) structure will not natively embed onto the available qubit architecture on the quantum chip. Thus, in these cases, a minor embedding of the problem structure onto the device is necessary. However, minor embeddings on these devices do not always make use of the full sparse chip hardware graph, and a large portion of the available qubits stay unused during quantum annealing. In this work, we embed a disjoint random QUBO on the unused parts of the chip alongside the QUBO to be solved, which acts as an indicator of the solution quality of the device over time. Using experiments on three different D-Wave quantum annealers, we demonstrate that (i) long term trends in solution quality exist on the D-Wave device, and (ii) the unused qubits can be used to measure the current level of noise of the quantum system.
\end{abstract}

\vspace{2pc}

\section{Introduction}
\label{sec:introduction}
Quantum annealing is a novel computing technology that uses quantum fluctuations to search for a global minimum of a combinatorial optimization problem \cite{PhysRevX.4.021041, morita2008mathematical, RevModPhys.80.1061, hauke2020perspectives, finnila1994quantum, kadowaki1998quantum, johnson2011quantum, PhysRevA.98.022314}. The quantum annealers manufactured by D-Wave Systems, Inc., are specialized hardware devices that implement quantum annealing. The class of problems that D-Wave quantum annealers can directly solve are defined by the function
\begin{equation}
    H(x_1,\ldots,x_n) = \sum_{i=1}^n h_i x_i + \sum_{i<j} J_{ij} x_i x_j,
    \label{eq:hamiltonian}
\end{equation}
where the linear weights $h_i \in \mathbb{R}$, $i \in \{1,\ldots,n\}$ and the quadratic couplers $J_{ij} \in \mathbb{R}$ for $i<j$ define the problem being solved. The task is to find a combination of the unknown binary variables $x_i$, $i \in \{1,\ldots,n\}$, that minimizes eq.~(\ref{eq:hamiltonian}). The function of eq.~(\ref{eq:hamiltonian}) is called a \textit{QUBO problem} if $x_i \in \{0,1\}$, and an \textit{Ising problem} if $x_i \in \{-1,+1\}$, where $i \in \{1,\ldots,n\}$. Both the QUBO and Ising formulations are equivalent \cite{Chapuis2017}. In this paper, we consider the QUBO version.

\begin{figure}
    \centering
    \includegraphics[width=0.2\textwidth]{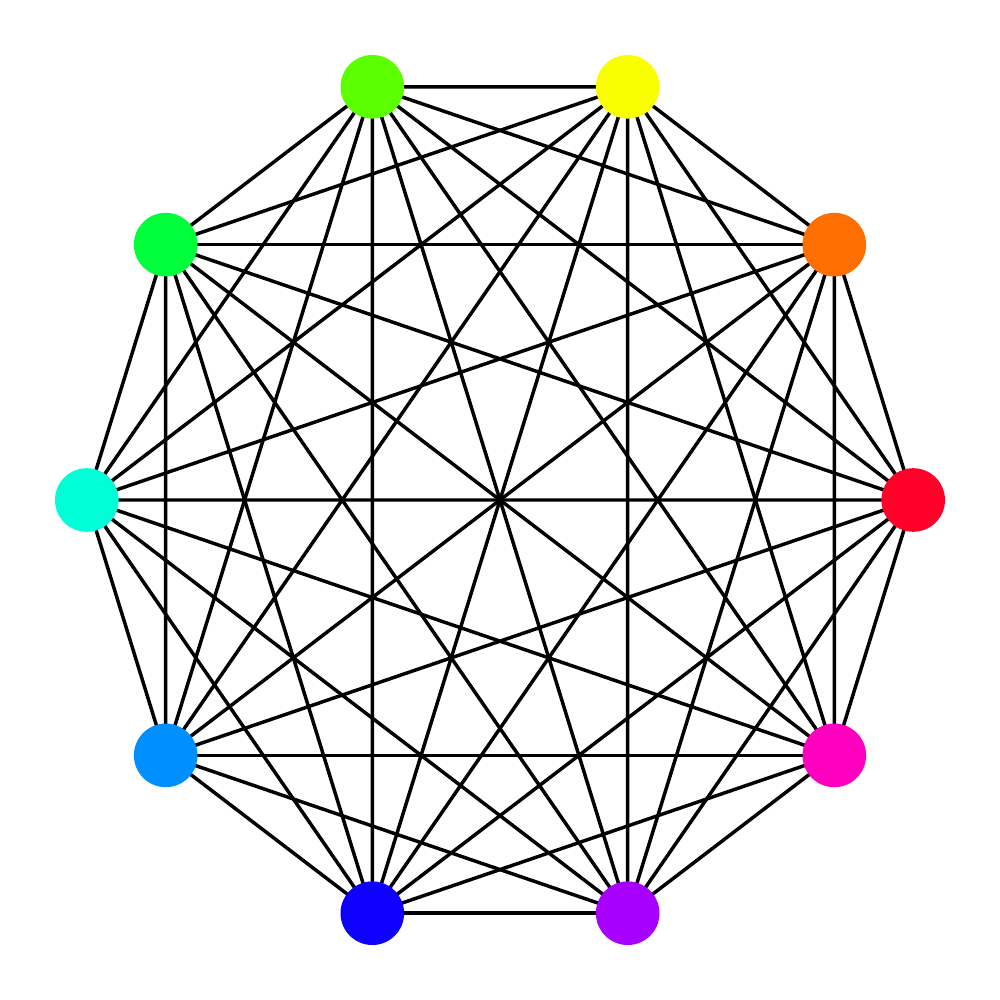}\\
    \includegraphics[width=0.3\textwidth]{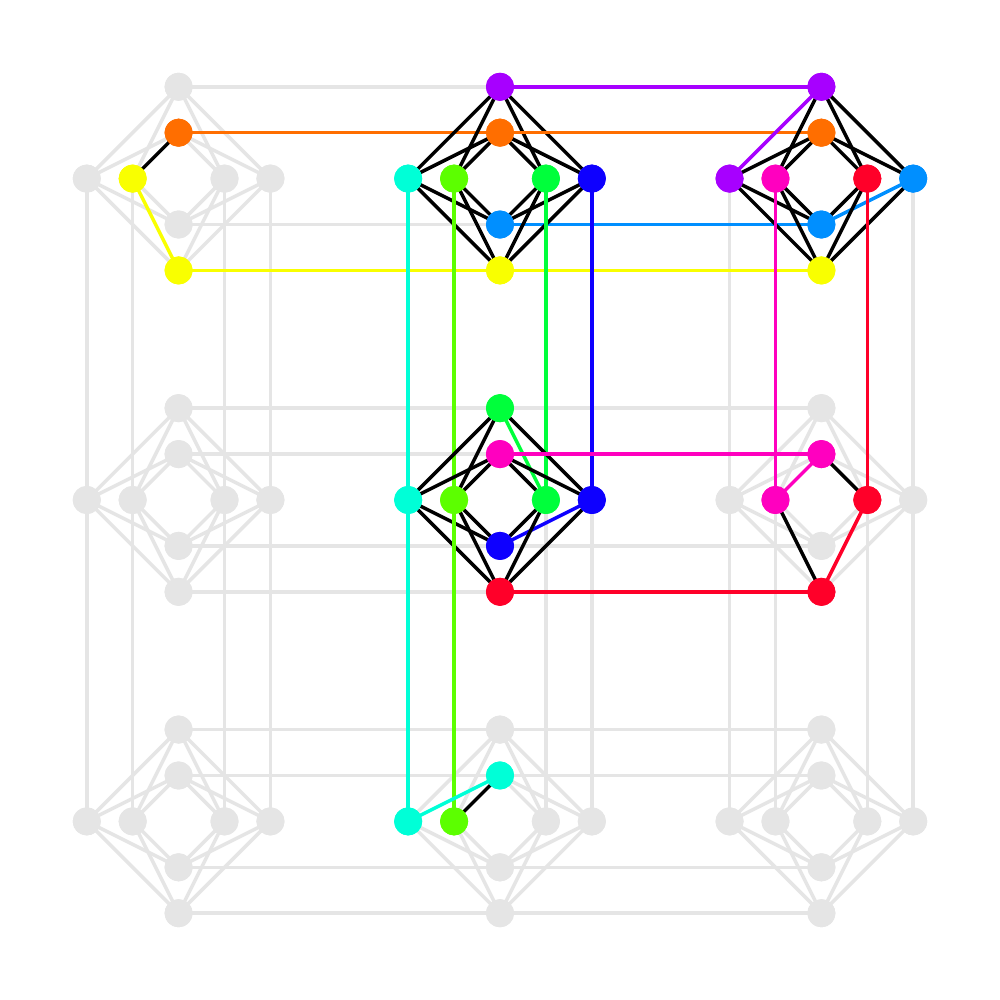}
    \includegraphics[width=0.3\textwidth]{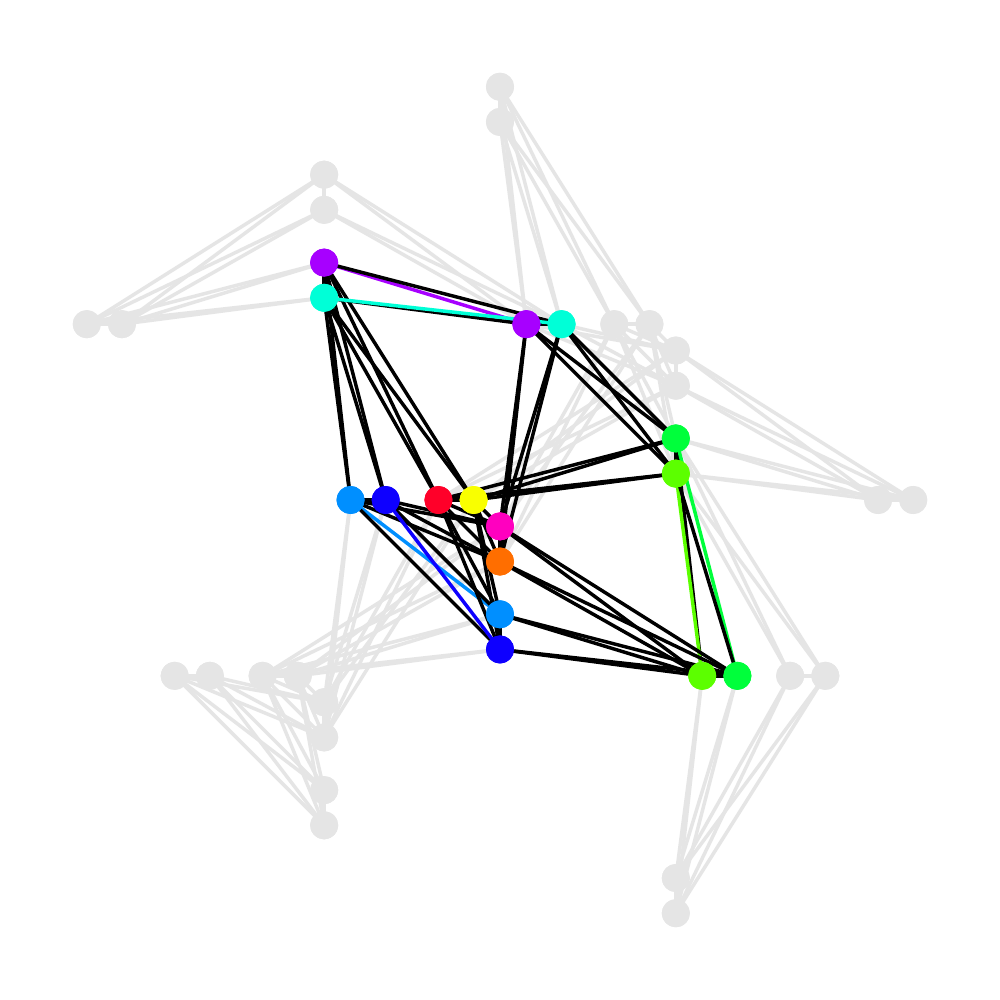}
    \caption{Example of a heuristic minor-embeddings of a $10$ node clique (top) onto a Chimera graph $C3$ (bottom left) and a Pegasus $P2$ graph (bottom right). Corresponding logical and physical qubits in each example share the same color. Note that the largest clique that can be minor embedded onto a Chimera $C3$ graph has size $12$, and the largest clique that can be minor embedded onto a Pegasus $P2$ graph is $10$. }
    \label{fig:example}
\end{figure}

Before attempting to minimize a function of the form of eq.~(\ref{eq:hamiltonian}) using a D-Wave quantum annealer, the functional form of eq.~(\ref{eq:hamiltonian}) must be mapped onto the D-Wave quantum chip. However, the connectivity structure of modern quantum annealing hardware is relatively sparse, and typically does not match the one of the problem being solved. Therefore, it is necessary to compute a \emph{minor embedding} of the graph representing the logical qubit structure of eq.~(\ref{eq:hamiltonian}) onto the graph representing the hardware qubit structure of the D-Wave quantum chip, where a connected set of hardware qubits called a \textit{chain} is used to represent a single logical qubit. An example can be found in Figure~\ref{fig:example}. Although the problem of finding a minor embedding is in general NP-hard itself, feasible heuristics have been developed for practical use \cite{Cai2014, https://doi.org/10.48550/arxiv.1507.04774, lobe2021minor, Zbinden2020}. Instead of computing a new, tailored minor embedding for each problem being implemented onto the D-Wave quantum chip, a fixed embedding of a complete graph of order $n$ can be computed and used for any problem of up to $n$ variables. This saves the time for computing the embedding, which often dominates the total time to find a solution of the original problem. However, regardless of whether a problem-tailored or a fixed embedding is used, a considerable proportion of hardware qubits are typically left unused depending on the size of the minor embedding and the connectivity of the quantum annealer.

After minor-embedding a given NP-hard problem of the type of eq.~(\ref{eq:hamiltonian}), a solution can be constructed via a quantum annealing process. However, due to the limitations of the current intermediate-scale quantum (NISQ) technology \cite{Preskill2018quantumcomputingin}, the solution quality might not be very high as it depends on factors such as decoherence from environmental impact, calibration errors, leakage between adjacent qubits, control errors, and fluctuations of the effective temperature. Hence, the amount of noise and the quality of the solutions can vary significantly during the operation of a quantum annealer. The aim of this work is to show that (i) the amount of noise is not completely random but follows longer term up and down trends affecting the solution quality on the D-Wave device, and (ii) that the idle qubits can be used to provide information on the current noise level and performance of the device.

To measure and predict the impact of various factors on the quality of the solution, which we refer to collectively as \textit{noise} for simplicity, we exploit the unused hardware qubits on the D-Wave chip as follows. We plant a randomly generated QUBO on the unused qubits, which then acts as an indicator of the error rate experienced by the quantum device. Specifically, we propose to divide up the hardware qubits on the quantum annealing chip into two sets. First, an all-to-all minor embedding of a complete graph of certain size $n$ is computed, which is kept fixed afterwards and utilized to embed different problems of size $n$ or less being solved. Second, a random QUBO is being generated on the remaining unused hardware qubits and the couplers between them. This QUBO is independent of the problem of interest being solved. We call that QUBO a \textit{performance indicator}. Since the performance indicator does not change and is being re-solved again with each run of the system, in ideal conditions and hardware one should always get solutions of equal quality. Hence, any significant changes in the quality of its solution can be interpreted as changes of the noise level of the D-Wave hardware. Our hypothesis is that, by analyzing the results from the second set of qubits (the performance indicator), we can obtain information about the quality of the solution on the first set of qubits (the problem QUBO).

All results presented in this work have been computed on three different D-Wave devices: The D-Wave 2000Q computer located at Los Alamos National Laboratory (LANL), referred to by its chip ID \texttt{DW\_2000Q\_LANL} in the remainder of this article, the D-Wave 2000Q device accessed through D-Wave Leap with the chip ID \texttt{DW\_2000Q\_6}, and the D-Wave Advantage System~4.1 accessed through D-Wave Leap, which we refer to by its chip ID \texttt{Advantage\_system4.1}.

The contributions of this article are fourfold:
\begin{enumerate}
    \item We establish that the noise on the chip follows long-term as well as short-term up and down trends.
    \item We demonstrate that the idle qubits can can be put to use, namely, in the design of a performance indicator to monitor the accuracy performance of the quantum chip.
    \item We show that there exists significant correlation between the measured energies of the samples of the problem QUBO and the performance indicator.
    \item We propose a simple algorithm to make use of the performance indicator, which observes the performance indicator over a certain burn-in period, and then samples solutions for the problem being solved whenever the performance indicator shows that the hardware is in a low-noise state. We demonstrate that the solutions thus obtained are considerably better than samples obtained from D-Wave in the conventional way (that is, at arbitrary times).
\end{enumerate}

The Python code and data associated with the results shown in this article are available in a Github repository\footnote{\url{https://github.com/lanl/Noise-Indicator-QA}}.

The article is structured as follows. After a literature review in Section~\ref{sec:literature}, we formalize the idea of estimating noise with the help of idle qubits in Section~\ref{sec:methods}. Section~\ref{sec:results} presents our experimental results, particularly on quantifying the noise on a D-Wave annealer with the help of the performance indicator, and on a simple technique to improve the solution quality on the D-Wave annealer. The article concludes with a discussion in Section~\ref{sec:discussion}.

\subsection{Literature Review}
\label{sec:literature}
Several recent works available in the literature attempt to characterize the noise of a quantum annealer, including the characterization of single qubit fidelity.

In \cite{Zaborniak2020,Zaborniak2021} the authors are interested in characterizing fluctuations of the parameters defining the original problem Ising caused by various noise sources on the chip. They introduce a method to benchmark the amount of noise affecting the programmed Ising of a quantum annealer and introduce an estimate of the noise spectral density affecting the problem parameters.

In \cite{ayanzadeh2021multi} a post-processing method called multi-qubit correction is proposed, which mitigates errors of quantum annealing samples by detecting and correcting local groups of qubits, which were read out in a excited state.

In \cite{1503.05679} a calibration procedure is developed to detect and correct persistent biases on two D-Wave Two devices. In \cite{Nelson2021} the authors introduce the QASA (Quantum Annealing Single-qubit Assessment) protocol, which computes relevant qubit performance metrics, analogous to gate model circuit operation error (or fidelity) rates \cite{alexander2020qiskit, sanders2015bounding}, across the entire quantum annealing device for a given quantum annealing parameter combination (the parameter set includes annealing time, the anneal schedule, etc). The QASA protocol is demonstrated on a D-Wave 2000Q device. Later, a similar protocol called Q-RBPN is proposed, which provides performance information of single qubits across both quantum annealing and gate model devices \cite{https://doi.org/10.48550/arxiv.2108.11334}.

In a review paper \cite{Yarkoni2022}, the authors remark that despite finite temperature effects present in samples coming from a quantum annealing processor, such samples are still useful in practice \cite{Raymond2016}. Moreover, even in the absence of quantum coherence and entanglement, in practice, quantum annealers still perform useful heuristic optimization \cite{Andriyash2017,Denchev2016}.

In \cite{Pudenz2014, PhysRevA.91.042302} an algorithm called Quantum Annealing Correction (QAC) is proposed, which provides a substantial improvement in performance compared to linear chains with no QAC; substantial analysis and improved variants have subsequently been implemented \cite{Pearson2019, Vinci2016, PhysRevLett.116.220501, vinci2015quantum}. QAC works by encoding a problem using a repetition code in conjunction with a bit-flip penalizing Hamiltonian. In a similar fashion, in \cite{Suzuki2020} the authors propose a method to suppress noise on the D-Wave quantum annealer by modifying the QUBO to be implemented on the D-Wave hardware.

Investigations regarding noise in gate model quantum computers are also available in the literature. For instance, in \cite{Harper2020} the authors introduce a protocol to estimate the effective noise and to detect correlations within arbitrary sets of qubits. The protocol is showcased on a 14-qubit superconducting quantum architecture. Similarly, in \cite{Shaib2021} the authors model the multi-qubit average behavior of a quantum system with a special set of quantum channels and gates, with the aim to quantify state preparation and measurement errors. In \cite{proctor2020detecting}, protocols are developed for tracking error, and therefore calibration, drift of circuit model quantum computers over time. A broad overview of quantum error correction and fault-tolerant computation can be found in \cite{Devitt2013}, covering coherent quantum errors and decoherence, simple quantum error correction codes, stabilizer codes, and fault-tolerant quantum error correction. Another introductory guide to the theory and implementation of quantum error correction codes can be found in \cite{Roffe2019}. An experimental evaluation of noise characterization and error mitigation on (universal gate) IBM Quantum computers can be found in \cite{9283531}. Error correcting code implementations are evaluated on IBMQ gate model quantum computers in \cite{Ahsan_2022, ahsan2020reconfiguring}. In \cite{Hamilton2020}, a matrix-based characterization method of qubit correlations called the measurement fidelity matrix (MFM) in gate model quantum computers is introduced. Additionally, in \cite{Dasgupta2021}, the authors propose to monitor multiple metrics to characterize both noise of quantum devices and fluctuations in device parameters.

Finally, there are alternative methods for using potentially idle qubits in modern sparsely connected quantum annealing devices when the size of the problem being solved is much smaller than the maximum size of a complete graph that can be embedded. For example, \emph{parallel quantum annealing}, also known as \emph{tiling} \footnote{\url{https://dwave-systemdocs.readthedocs.io/en/samplers/reference/composites/tiling.html}}, has been experimentally evaluated before \cite{pelofske2022quantum, pelofske2022parallel, https://doi.org/10.48550/arxiv.2205.12165}. Similarly, the sparse connectivity on NISQ circuit model quantum computers has been used to execute circuits in parallel \cite{https://doi.org/10.48550/arxiv.2209.03796, https://doi.org/10.48550/arxiv.2102.05321, 9407180, 10.1145/3352460.3358287, https://doi.org/10.48550/arxiv.2112.00387, 9749894}.

\section{Methods}
\label{sec:methods}
The idea of this work is to establish the existence of trends in the level of noise  of the D-Wave device and then use idle qubits on the D-Wave quantum chip to characterize such noise while solving optimization problems simultaneously.

The optimization problem we would like to solve (referred to as \textit{problem QUBO}) can be any, as long as it fits onto an embedding of a complete graph of a certain size (either the largest one that fits onto the D-Wave chip, or smaller), which is held fixed during the experiments. Our main innovation is to read out the energies of a randomly generated QUBO (called the \textit{performance indicator}) occupying some or all of the remaining hardware qubits on the D-Wave chip that are left after embedding the problem QUBO. To this end, we establish that the energies observed for the performance indicator and the problem QUBO being solved are positively correlated. Therefore, monitoring the solution quality of the performance indicator allows a user to determine if the solutions found for the problem QUBO of interest are below or above the average quality achievable by the annealer.

The following subsections elaborate on our methodology, in particular:
\begin{enumerate}[leftmargin=4em]
    \item choosing the idle qubits to monitor (Section~\ref{sec:idle});
    \item computing an embeddding for the problem QUBO and, optionally, the performance indicator (Section~\ref{sec:embeddings});
    \item the actual specification of the weights of the performance indicator (Section~\ref{sec:indicator});
    \item comparing the energy reads for both the problem QUBO and the performance indicator (Section~\ref{sec:comparison});
    \item the algorithm we use to utilize the performance indicator (Section~\ref{sec:algo}).
\end{enumerate}

\subsection{Identifying idle qubits on the D-Wave quantum chip}
\label{sec:idle}
We start by fixing two sets of hardware qubits on the hardware connectivity graph. These sets can have arbitrary sizes, and not all hardware qubits need to be used. Note that there are many different and valid possibilities of implementing the idea of a problem QUBO and an performance indicator to monitor D-Wave noise.

In this paper, we utilize the following approach. We first compute a minor embedding of a complete graph onto the hardware connectivity graph (also called a \textit{clique embedding}), which will be used to embed the problem QUBO. The size of the clique embedding we use varies across our experiments. The clique embedding allow one to embed the logical qubits onto the physical hardware by creating ferromagnetic chains of qubits, which represent logical variable states. After embedding the problem QUBO, we set the performance indicator to be the induced subgraph of all qubits (nodes) that were not used in the minor embedding. Thus, it is possible that the performance indicator itself is not connected.

Alternatively, one can also fix two embeddings (possibly two clique embeddings) for both the problem QUBO and the performance indicator that do not occupy all available hardware qubits. This approach allows one to analyze the behavior of the performance indicator, and its correlation to the problem QUBO, in the scenario where the two QUBOs are comparable in size but small in comparison to the hardware graph.

Extensions of this idea are possible, for instance by employing three or more clique embeddings to solve several problems simultaneously while having some hardware qubits act as the performance indicator.

\subsection{Embeddings for both the problem QUBO and performance indicator}
\label{sec:embeddings}
\begin{figure}[h]
    \centering
    \includegraphics[width=0.4\textwidth]{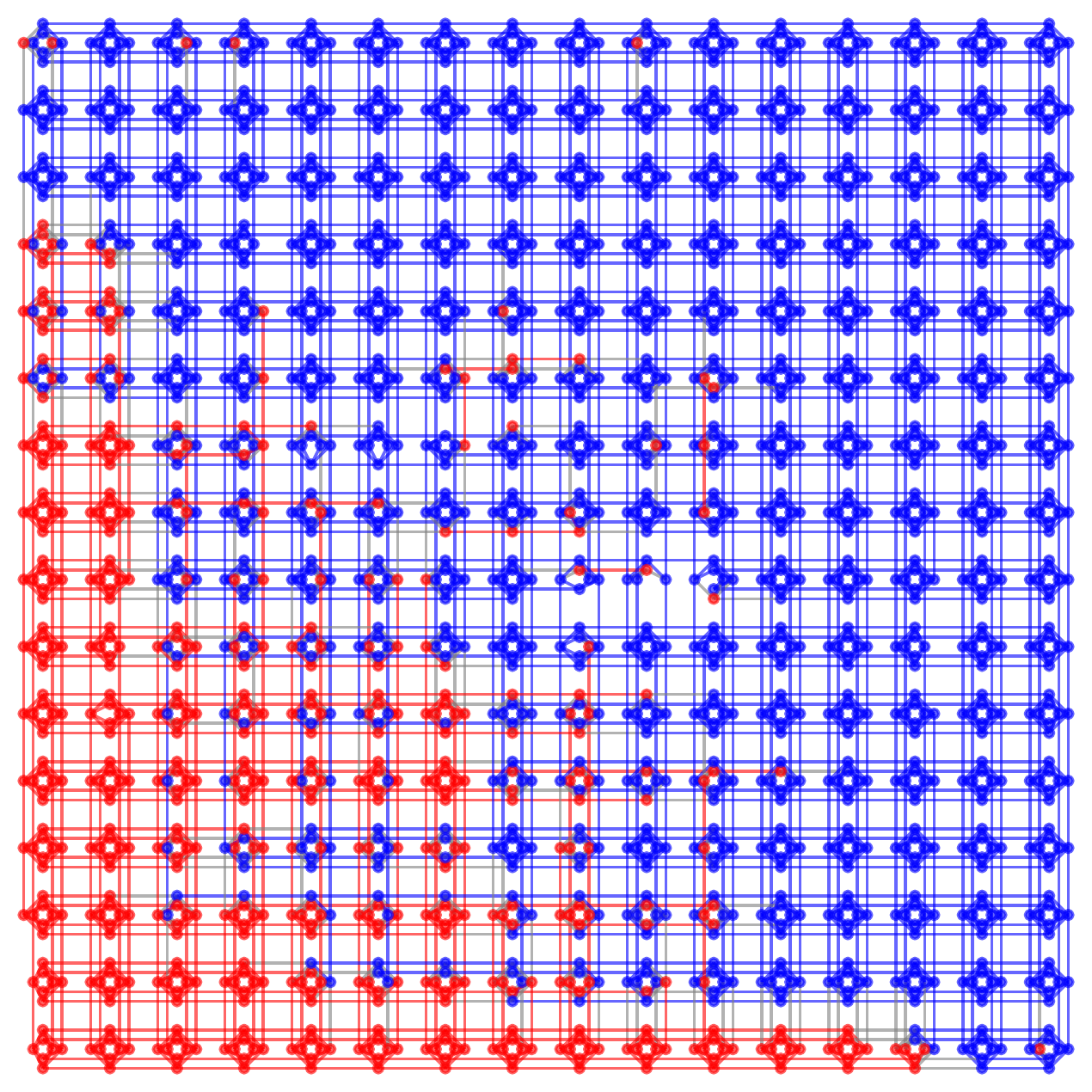}
    \includegraphics[width=0.4\textwidth]{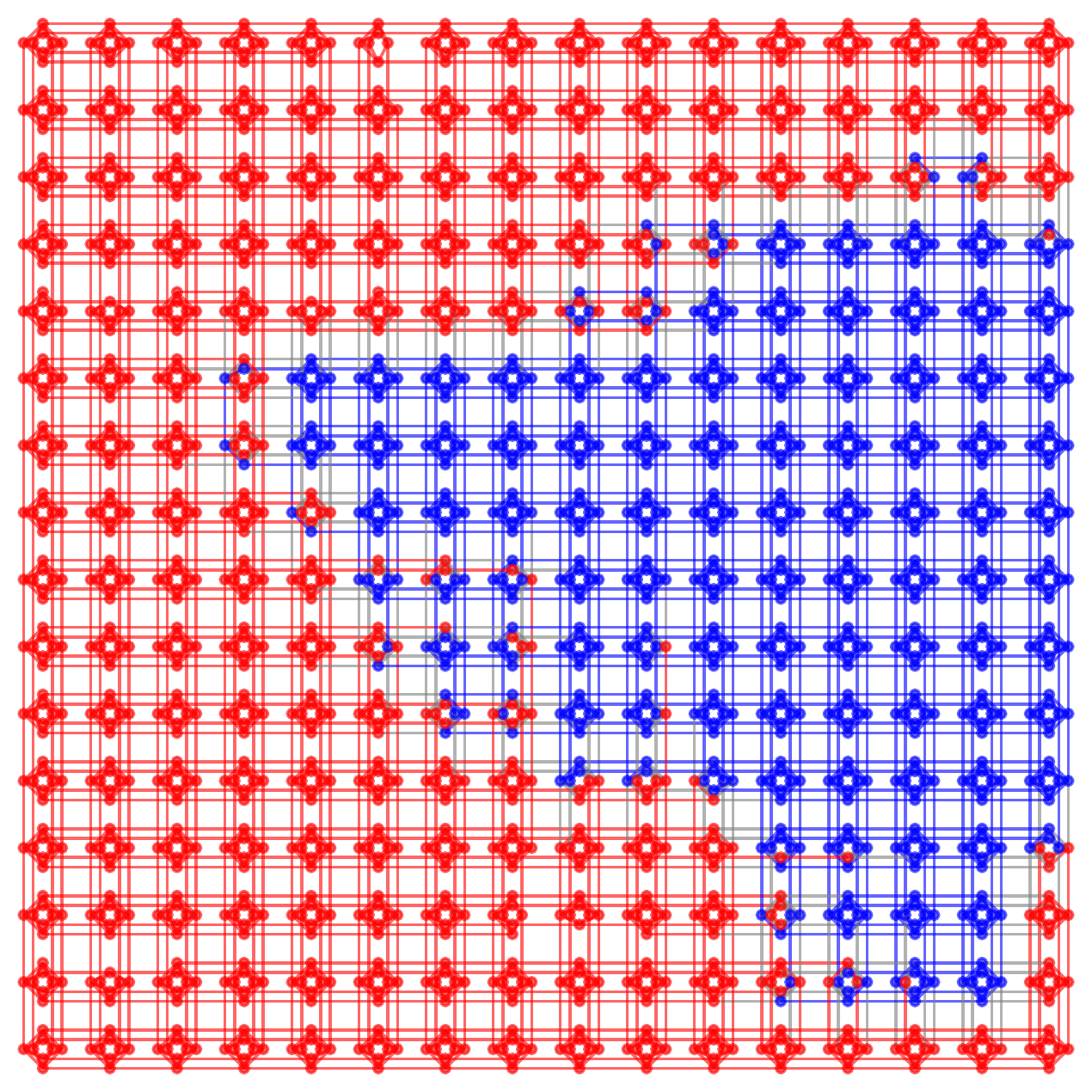}\\
    \includegraphics[width=0.4\textwidth]{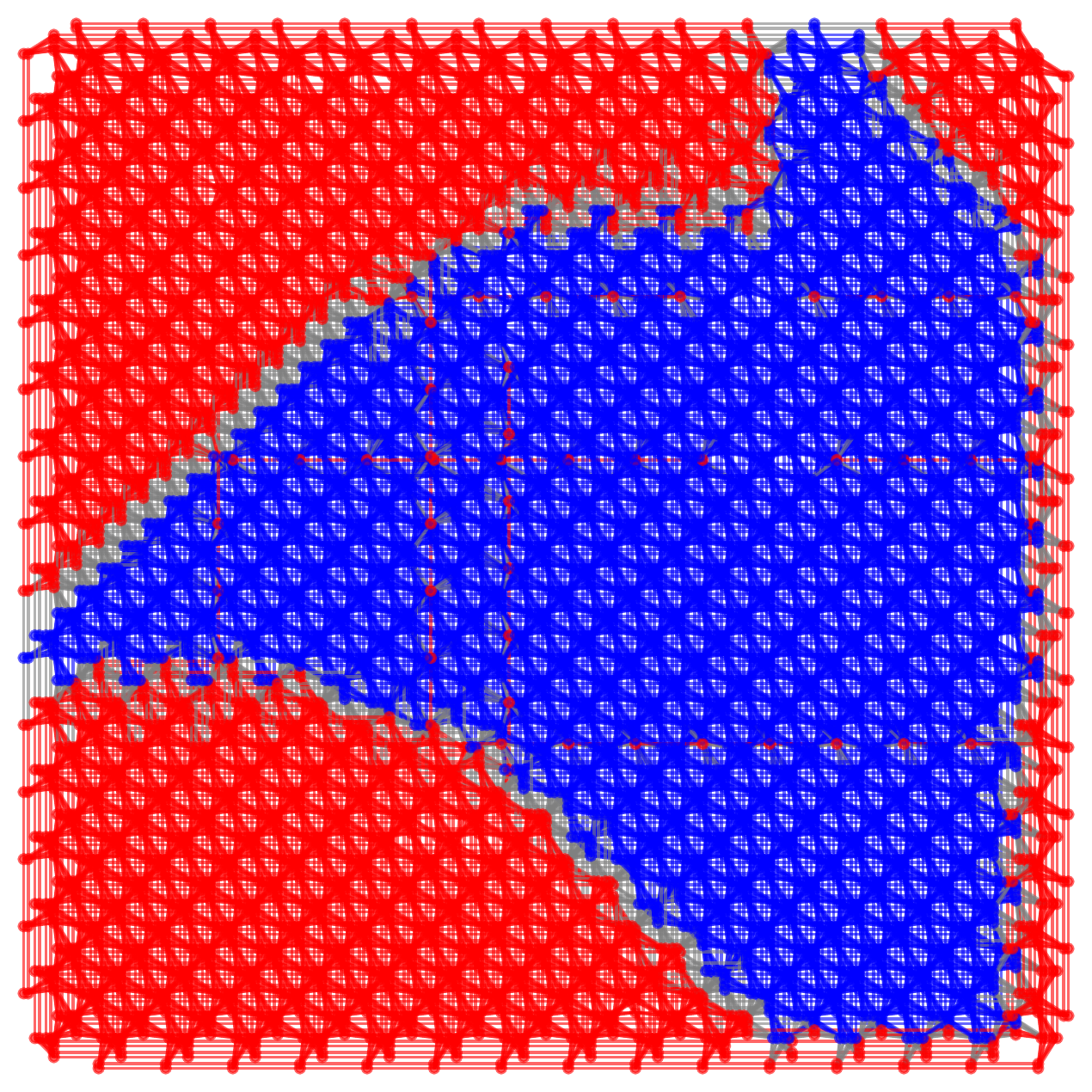}
    \includegraphics[width=0.4\textwidth]{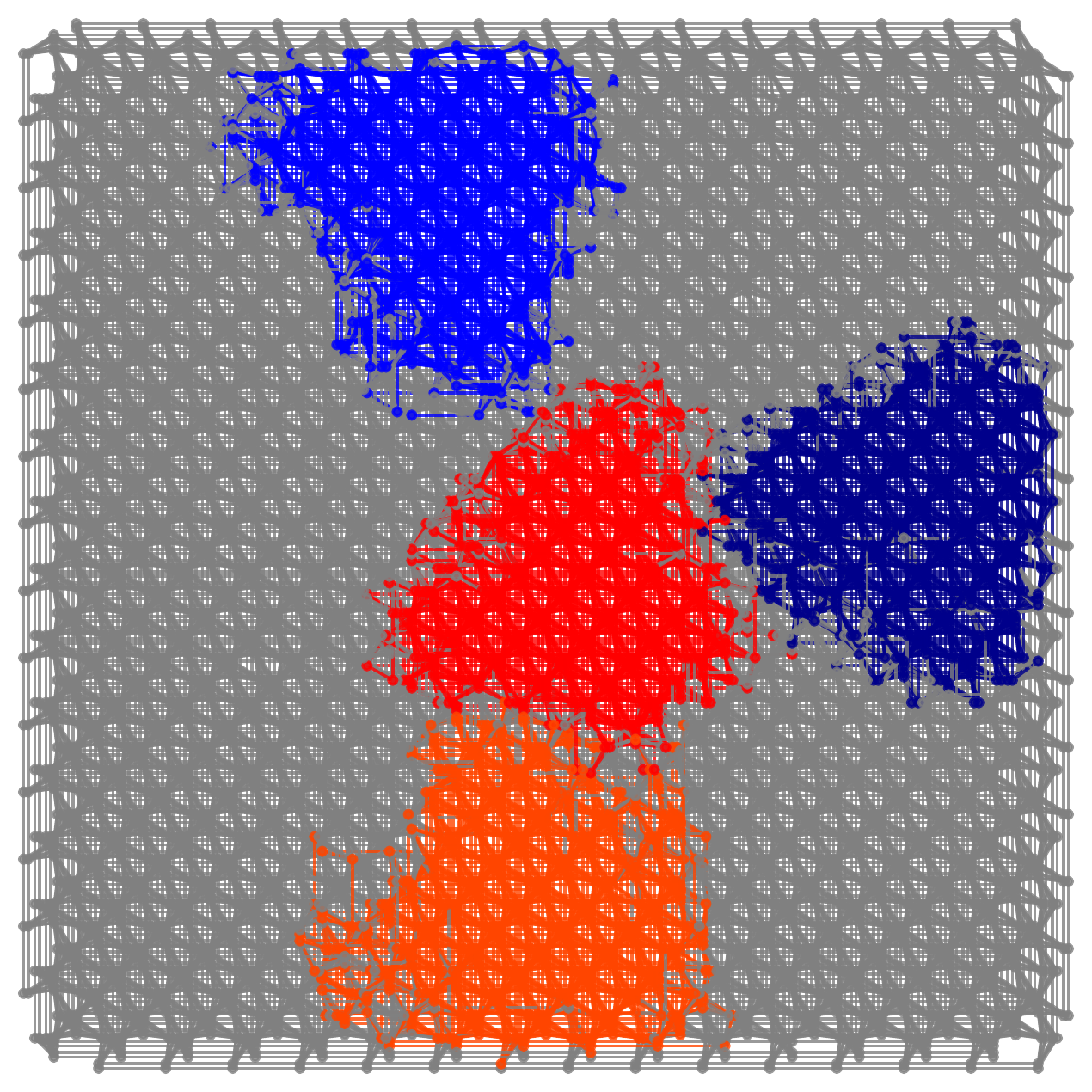}
    \caption{Clique embeddings used in the experiments. Blue color encodes the clique embedding used by the problem QUBO and red color encodes the remaining region onto which the performance indicator is embedded. LANL D-Wave 2000Q (top left) with complete 65 node embedding, D-Wave 2000Q 6 (top right) with complete 50 node embedding, D-Wave Advantage System~4.1 (bottom row) with a complete 80 node embedding (bottom left), and four parallel embeddings of size 60 not occupying the entire chip (bottom right). The figures show all edges between the nodes in each complete embedding for illustration purposes, even though not all of them might be used. These renderings represent the exact QA chip connectivity, which means that any defective qubits in the hardware will also not be present in the diagrams.}
    \label{fig:embeddings}
\end{figure}
Figure~\ref{fig:embeddings} shows an example of the embeddings used in our experiments. The clique embedding for the problem QUBO is always colored in blue, while the remaining qubits used for the performance indicator are colored in red. Many different topologies are possible. For instance, the performance indicator can be smaller (Figure~\ref{fig:embeddings}, top left) or larger (top right) than the embedding of the problem QUBO. Moreover, the two embeddings do not have to occupy all available hardware qubits, allowing one to spatially separate the active qubits being used (bottom right). We do not use any couplers between hardware qubits connecting the problem QUBO and performance indicator, thus causing the minor embedded problem QUBO and the performance indicator to be disjoint.

All minor embeddings used in this work were computed using the function \textit{find\_embedding} of the D-Wave Ocean SDK \cite{ocean} (specifically the method \textit{minorminer} of \cite{Cai2014}). We utilized this function with the parameters \textit{max\_no\_improvement} set to 300, \textit{chainlength\_patience} set to 300, and \textit{tries} set to 300 \cite{Cai2014}. The only exception is the $177$ clique embedding for the Advantage\_system4.1, where we used the precomputed clique provided in the D-Wave Ocean SDK. Moreover, the linear coefficients are always uniformly distributed across all chained qubits, and the quadratic terms are uniformly distributed over the available physical couplers between chains.

\subsection{The choice of the performance indicator}
\label{sec:indicator}
Once the hardware qubits for the performance indicator are fixed, one needs to specify the actual QUBO weights of the performance indicator. This is necessary as any QUBO being implemented on the chip alongside the actual QUBO to be solved has to be fully determined. With respect to the types of weights of the indicator, we employ the following two specific performance indicator types in our experiments:
\begin{enumerate}[leftmargin=4em]
    \item[PI1] For each linear and quadratic weight, we randomly generate a weight by sampling from a uniform distribution in $(-1, 1)$.
    \item[PI2] For each linear and quadratic weight, we randomly generate a weight by assigning either $-1$ or $1$ with probability $0.5$.
\end{enumerate}
Both QUBOs obtained in this way are spin glass models with random weights and, therefore, in general NP-hard \cite{Barahona1982,Lucas2014}. In particular, we do not know the ground state configuration minimizing each performance indicator. However, this is not necessary, as samples from any D-Wave device obtained for the performance indicator are solely used for monitoring the state of the annealer, and after having observed sufficiently many samples it will be possible to determine the range of energies for the problem QUBO with high confidence.

The last issue to address when solving the two problems (the actual problem QUBO $Q_P$ and the performance indicator $Q_I$) simultaneously on the D-Wave quantum device pertains to their normalization. Typically, the range of weights used in both QUBOs is not identical. While the absolute values of the coefficients of both performance indicators are in $[-1,1]$ by construction, those of the problem QUBO can be as high as the chain strength that is used to embed the problem. D-Wave uses a function called \textit{auto\_scale} to multiply all coefficients by the same constant in order to ensure that the quadratic coefficients are in the range $[-2,2]$, and the linear ones in $[-1,1]$. That means that autoscaling could make some coefficients of the performance indicator too small, affecting the accuracy of its representation and the quality of the annealing result. Therefore, when implementing the combined QUBO $Q$, we multiply each weight of the performance indicator by a constant $C$, allowing us to control how both QUBOs are relatively weighted when applying the \textit{auto\_scale} option of the D-Wave devices. 
This constant is set to $C = |Q_P|/|Q_I|$, where $|Q_P|$ and $|Q_I$ denote the maximum coefficient in absolute value of the QUBOs $Q_P$ and $Q_I$, respectively, resulting in a combined QUBO $Q=Q_P+|Q_P|/|Q_I| \cdot Q_I$.

\subsection{Comparison with the problem QUBO}
\label{sec:comparison}
The purpose of implementing the performance indicator alongside the actual QUBO being solved on the D-Wave chip is to monitor the quality of the D-Wave annealing results. In particular, we aim to infer conclusions about the quality of the samples returned for the problem QUBO from the ones observed for the performance indicator.

We use the following procedure to determine the similarity between the energy time series for the problem QUBO and for the performance indicator. First, we normalize the energies observed for each QUBO separately by scaling all values into the range $[0, 1]$. Afterwards, we adjust the time series for the performance indicator in such a way that its mean matches the one of the time series of the problem QUBO. Normalization is needed because the problem QUBO and performance indicator, having different sizes and coefficients, normally take values in different ranges. Since we aim to compare trends, not values, we set $x'_i = x_i - (m_x-m_y)$, where $x=\{x_1,\ldots,x_N\}$ and $y=\{y_1,\ldots,y_N\}$ are the time series of length $N \in \N$ observed for the problem QUBO and the performance indicator, respectively, and $m_x$ and $m_y$ are their means.

We measure the similarity of the two observed time series with the help of the root-mean-square deviation (RMSD) of the difference between $x$ and $y$, given as

\begin{equation}
	\mathrm{RMSD}(x,y) = \sqrt{\frac{1}{N}\sum_{i=1}^N (x_i - y_i)^2}.
	\label{eq:rmsd}
\end{equation}

\subsection{A two-phase procedure to leverage the performance indicator}
\label{sec:algo}
After setting up the problem QUBO and the performance indicator on the D-Wave quantum device, we can make use of the additional information provided by the performance indicator in various ways. For instance, we can run the following two-phase procedure.

In a \textit{set-up phase}, we obtain samples for the problem QUBO and, more importantly, for the performance indicator. After a prespecified burn-in phase of length $b \in \N$ runs, we store a history and/or summary statistic of the values observed for the performance indicator, with the aim to reuse them in the future for solving problems from the same class.

In a second, \textit{usage phase}, we obtain samples from both the problem QUBO and the performance indicator simultaneously, and try to estimate the quality of the former using the values of the latter plus the stored statistics. For instance, one can compute the percentile rank of the performance indicator energy of the last run in comparison with the saved values. That percentile value can then accompany the problem QUBO result to inform about its estimated quality.

In a more complex scenario, if a sample coming from the performance indicator is less than a prespecified threshold $\tau \in (0,1)$, we accept the corresponding energy value of the problem QUBO (at the same time point) as an acceptable solution. This threshold is usually calibrated using the history of samples obtained for the performance indicator, and can be updated while the algorithm is running. Otherwise, we reject the sample returned for the problem QUBO as probably being of low quality and request another anneal at a later time.

\section{Results}
\label{sec:results}
In this section we present an experimental analysis of the performance indicator and its relationship with the problem QUBO.

We start by introducing the D-Wave hardware we employ (Section~\ref{sec:dwave}), the Maximum Clique and Minimum Vertex Cover problems under consideration (Section~\ref{sec:maxclique}), and our experimental setting (Section~\ref{sec:setting}).

We establish that the D-Wave annealers exhibit long term trends in the solution quality of the samples they return (Section~\ref{sec:trends}). We then demonstrate that the reads from the performance indicator are indeed correlated with the ones of the problem QUBO for all three D-Wave annealer systems under consideration (Section~\ref{sec:correlations}), and investigate the stability of the performance indicator across problem instances (Section~\ref{sec:alternating}). We conclude by showing that the samples returned from Advantage\_system4.1 at times when the performance indicator suggests that the annealer is in a state of low noise are considerably better than those obtained the conventional way (Section~\ref{sec:histograms}). A summary of all time series experiments can be found in~\ref{sec:appendix}.

\subsection{The D-Wave Quantum Annealing hardware}
\label{sec:dwave}
As mentioned in Section~\ref{sec:introduction}, we use three different D-Wave quantum annealers in this work, the \texttt{DW\_2000Q\_LANL}, the \texttt{DW\_2000Q\_6}, and the \texttt{Advantage\_system4.1}. The main difference between these devices is their number of available hardware qubits and the connectivity structure of those qubits. The \texttt{DW\_2000Q\_LANL} and \texttt{DW\_2000Q\_6} devices have roughly 2000 hardware qubits, connected in the \textit{Chimera} topology depicted in Figure~\ref{fig:embeddings} (top row), and the Advantage\_system4.1 device has roughly 5000 hardware qubits in a topology called \textit{Pegasus} depicted in Figure~\ref{fig:embeddings} (bottom row) \cite{Boothby2020, https://doi.org/10.48550/arxiv.1901.07636}.

The parameters of each D-Wave quantum annealer are kept at their default values in all experiments, with the exception of the following:
\begin{enumerate}
    \item The number of anneals is set to 100.
    \item The programming thermalization is chosen as 0 microseconds.
    \item The option to reduce intersample correlation is set to \textit{true}. This adds a slight time delay before each readout, thus reducing sample-to-sample correlation caused by the spin-bath polarization effect \cite{spin-bath-polarization, PhysRevApplied.15.014029}. 
\end{enumerate}
The readout thermalization parameter was always left to the default, which is 0 microseconds.

\subsection{The Maximum Clique and Minimum Vertex Cover problems}
\label{sec:maxclique}
This section defines the type of optimization problem that we solve on the quantum devices in our experiments.

We first look at the Maximum Clique (MC) problem. We are given an undirected graph $G=(V,E)$, where $V$ is a set of vertices and $E \subseteq V \times V$ is a set of edges. A subgraph $G(S)$ of $G$ induced by a subset $S \subseteq V$ is called a \textit{clique} of $G$ if $G(S)$ is complete, meaning that there exists an edge $(v,w) \in E$ for any $v,w \in S$, $v \neq w$. A \textit{maximum clique} of $G$ is a clique of $G$ of maximum size. The MC problem asks to find a maximum clique in $G$, and it is one of the most famous NP-hard problems with many applications in, for instance, bioinformatics, data mining, and network analysis.

In order to be able to solve an instance of the MC problem, we need to formulate it in a QUBO or Ising form as given in eq.~(\ref{eq:hamiltonian}). As shown in \cite{Chapuis2017,Pelofske2019mc}, a QUBO formulation for the MC problem on a graph $G=(V,E)$ is
\begin{equation}
    H_{MC} = -A\sum_{v \in V} x_v + B\sum_{(u,v) \in \overline{E}} x_u x_v,
    \label{eq:MC}
\end{equation}
where $\overline{E}$ denotes the edge set of the complement graph of $G$ and the constants $A>0$ and $B>0$ need to satisfy $A<B$. Without loss of generality, we define $A=1$ and $B=2$ in the remainder of this article. The binary variable $x_v \in \{0,1\}$ for each vertex $v \in V$ indicates if the vertex $v$ belongs to the maximum clique ($x_v=1$) or not ($x_v=0$).

Second, we consider the Minimum Vertex Cover (MVC) problem. For the undirected graph $G=(V,E)$ defined above, a subset $V' \subseteq V$ is called a \textit{vertex cover} if every edge in $E$ has at least one endpoint in $V'$, that is, if for every $e=(u,v) \in E$ it holds true that $u \in V'$ or $v \in V'$. A \textit{minimum vertex cover} is a vertex cover of minimum size. A QUBO formulation of MVC can be found in Section~4.3 of \cite{Lucas2014}, given by
\begin{equation}
    H_{MVC} = A \sum_{(u,v) \in E} (1-x_u)(1-x_v) + B \sum_{v \in V} x_v.
    \label{eq:MVC}
\end{equation}
Similarly to the encoding of the MC problem, each $x_v \in \{0,1\}$ for $v \in V$ in eq.~(\ref{eq:MVC}) is a binary variable indicating if vertex $v$ belongs to the MVC. The constants $A$ and $B$ have to satisfy $0<B<A$ as shown in \cite{Lucas2014}, and we fix $B=1$ and $A=2$ in the remainder of the article.

In our implementation, we used the uniform torque compensation feature for defining the QUBO with a \textit{UTC} prefactor of $1$ for \texttt{DW\_2000Q\_6} and Advantage\_system4.1, and a prefactor of $2$ for \texttt{DW\_2000Q\_LANL}. The uniform torque compensation method aims to reduce broken chains by computing the chain strength based on the square root of the mean of the quadratic coupler values of the QUBO \cite{uniform-torque-compensation}.

\subsection{Experimental setting}
\label{sec:setting}
In our experiments, the Python package \textit{NetworkX} \cite{networkx} was used to generate all random graphs, as well as to draw all connectivity graphs. The other drawings were generated using \textit{matplotlib} \cite{matplotlib, Hunter2007}. To construct minor embeddings onto the connectivity graph of the hardware qubits, we embed appropriately sized QUBOs for both the MC problem (see Section~\ref{sec:maxclique}) and for the performance indicator (see Section~\ref{sec:indicator}). The resulting minor embeddings for the three different D-Wave systems are shown in Figure~\ref{fig:embeddings}.

Next, we need to construct the problem QUBO and the performance indicator. The choice of the performance indicator (PI1 or PI2, see Section~\ref{sec:indicator}) is given individually for each experiment. The random graphs we consider to construct MC or MVC problem instances are Erd{\H o}s-R\'enyi random graphs with a graph density chosen uniformly at random in $(0,1)$ unless specified explicitly.

After embedding the QUBO for the MC problem onto the clique minor embedding, and the performance indicator onto the remaining hardware qubits of the QPU, we execute $60,000$ D-Wave calls for each problem, each having $100$ anneals. We use this data to generate time series for both the problem QUBO and the performance indicator by averaging the energies obtained in each batch of $100$ anneals, leading to two time series of length $60,000$.

We first look the raw data from all anneals. As this is noisy, a clearer view on the trends can be obtained by calculating a moving average, for which we always employ a window size of $500$. We measure the similarity between the reads for the problem QUBO and the performance indicator using the Pearson correlation coefficient and the root-mean-square deviation (RMSD), see Section~\ref{sec:comparison}. Here, a higher Person correlation and a lower RMSD indicate a better agreement of the reads for both QUBOs.

\subsection{Presence of long-term trends in the quantum annealer's performance}
\label{sec:trends}
Our first task is to show that a long-term trend in the quality performance of D-Wave indeed exists. Our experiment is based on the following idea: If we solve the same problem on D-Wave over and over, we should get (roughly) the same results, subject to random fluctuations (white noise). Specifically, one D-Wave call usually involves a large number of individual anneals (samples), up to $10,000$ for the current generations of the annealer, for efficiency reasons. Assume the returned samples are from a fixed distribution $D$ with mean $\mu$ and variance $\sigma^2$ (sampling from D-Wave is often modeled as sampling from a Boltzmann distribution). Then the average of all sample values returned from a D-Wave call should be normally distributed with mean $\mu$ and variance $\sigma^2/n$, where $n$ is the size of the sample, by the central limit theorem. That means that, if the distribution $D$ stays fixed and $n$ is reasonably large, the sample averages (with mean $\mu$ and variance close to zero for large $n$) should not change much across different anneals. Our experiments will show that this is not the case, i.e., that not only the averages vary significantly, but they go up or down in consistent, long-term patterns.

\begin{figure}[h]
    \centering
    \includegraphics[width=0.49\textwidth]{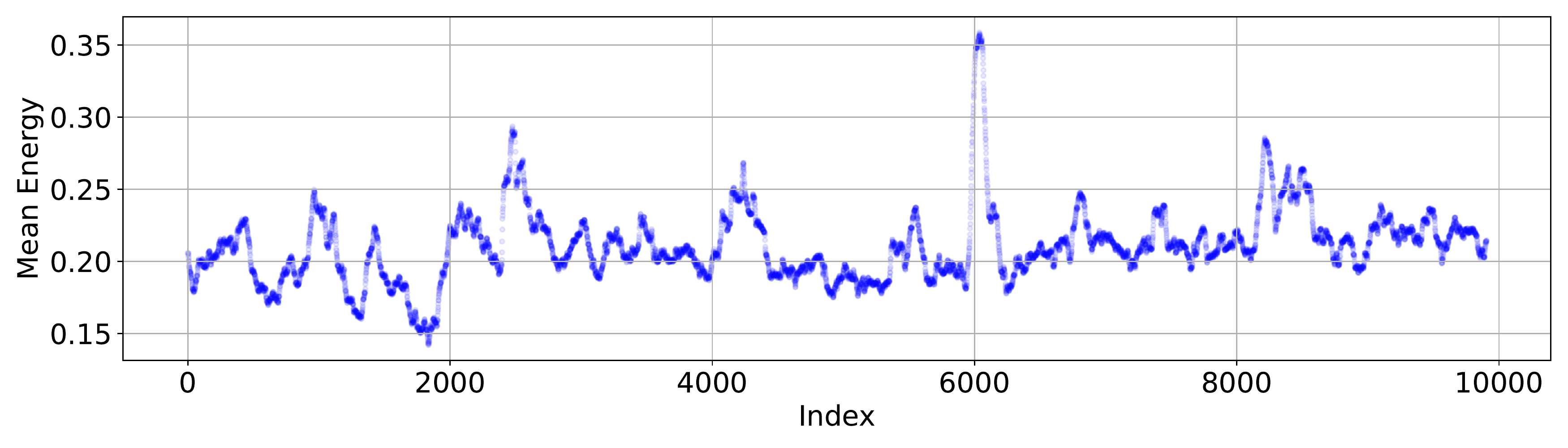}
    \includegraphics[width=0.49\textwidth]{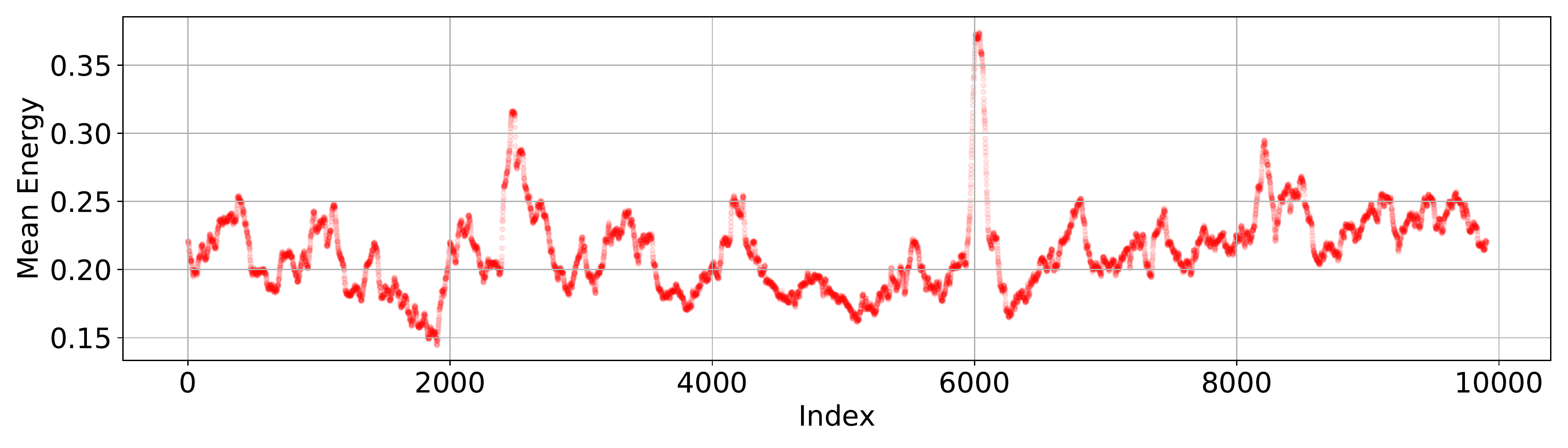}\\
    \includegraphics[width=0.49\textwidth]{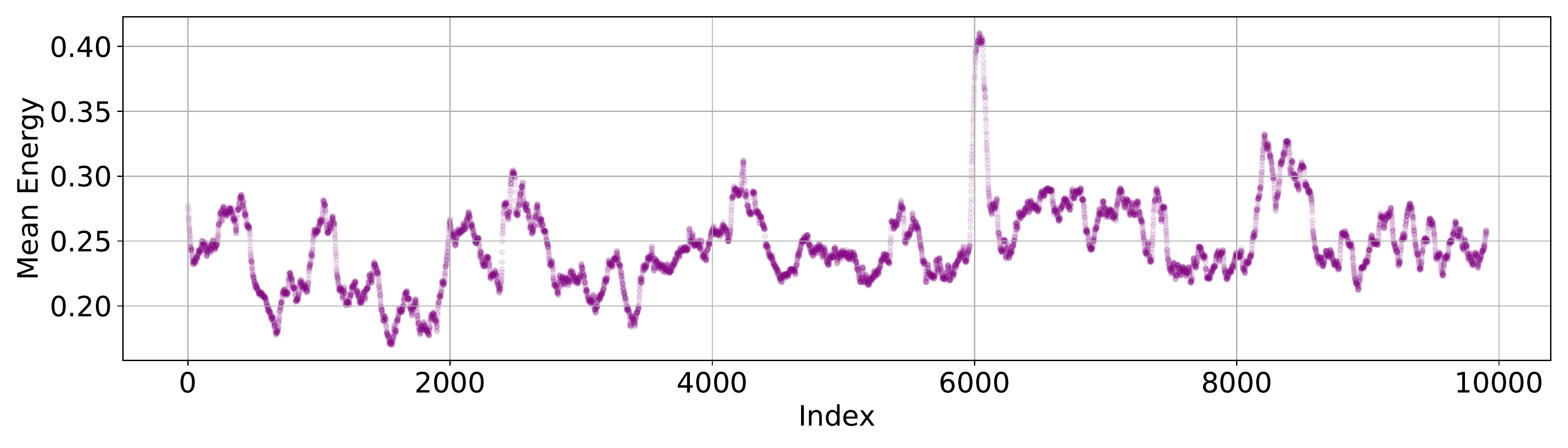}
    \includegraphics[width=0.49\textwidth]{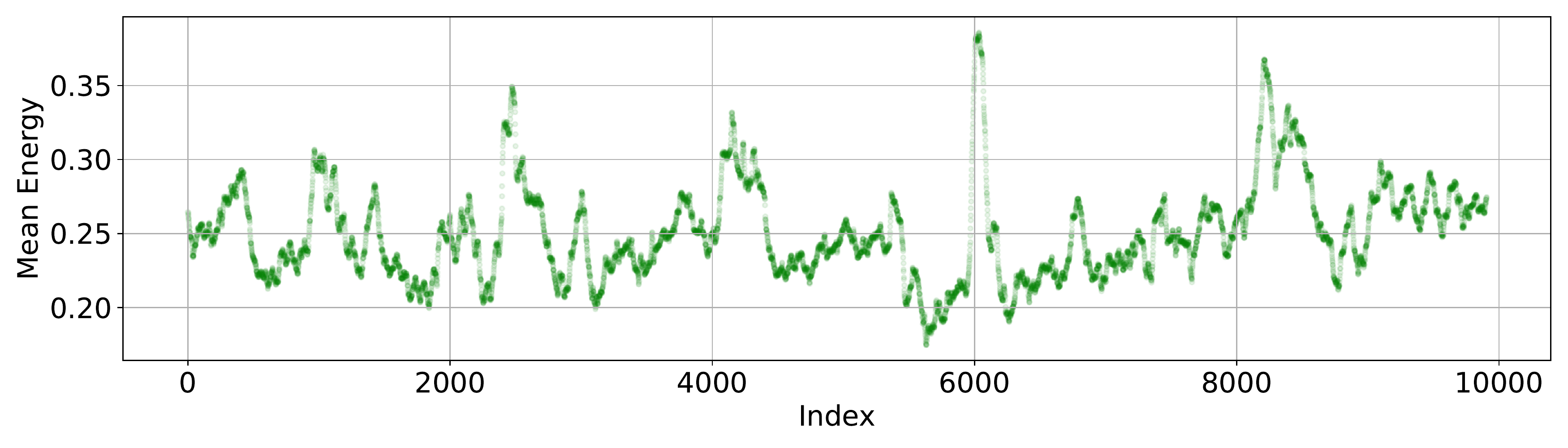}
    \caption{Longer term energy trends of four problems executed in parallel during the same annealing cycles using a total of 10,000 D-Wave calls. The data is plotting the mean energies per D-Wave call, with a moving average with window size 100. The top two plots are MC QUBOs of random graphs of size $60$ with graph density $0.3$ (left) and $0.7$ (right). The bottom two plots are MVC QUBOs of random graphs of size $60$ with graph density $0.3$ (left ) and $0.7$ (right). }
    \label{fig:four_problem_trends}
\end{figure}

To investigate the presence of trends in the time series data obtained for the D-Wave devices, we employ the four clique embeddings depicted in Figure~\ref{fig:embeddings} (bottom right), and embed four disjoint QUBOs on them. Those QUBOs represent four different optimization problems: two MC QUBOs and two MVC QUBOs, each on Erd{\H o}s-R\'enyi random graphs with $60$ vertices and graph densities of $0.3$ and $0.7$, respectively, and they are executed on the D-Wave QPU during the same annealing cycles (using the \textit{parallel quantum annealing} procedure from \cite{pelofske2022quantum, pelofske2022parallel}). All four problems were embedded using a chain strength of $5$. Moreover, due to their relatively similar size, their minimum and maximum QUBO weights were similar so normalization was not performed. A total of $10,000$ D-Wave calls were made, resulting in time series of length $10,000$.

\begin{figure}[h]
    \centering
    \includegraphics[width=0.89\textwidth]{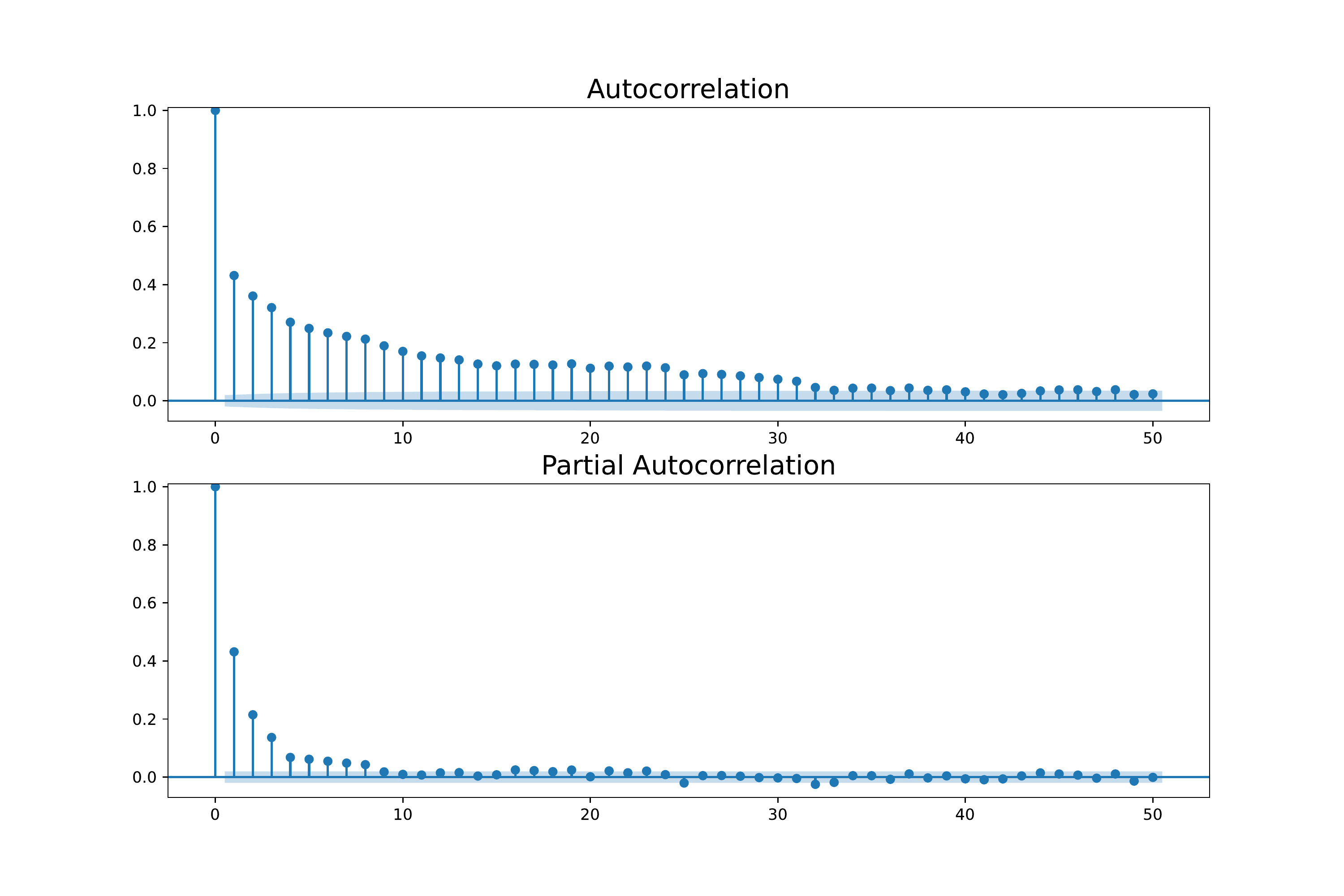}
    \caption{Auto correlation (top) and Partial auto correlation (bottom) for the time series of the mean energies from Figure~\ref{fig:four_problem_trends} (top left), computed on the raw mean energies (without moving average and without normalization of the mean). }
    \label{fig:four_problem_trends_ACF_PACF}
\end{figure}

Figure~\ref{fig:four_problem_trends} shows the obtained energy time series. Visually inspecting the plots, we can see that there appear to be partial up trends and down trends, and that they are shared among all four plots. Next, we aim to provide some \textit{statistical} evidence that the observed patterns in Figure~\ref{fig:four_problem_trends} can not simply be attributed to white noise. To this end, we consider the fit of an ARMA (autoregressive moving average) model, which is a linear model of stationary time series that aggregates two simpler models: the autoregression (AR) model and the moving average (MA) model. The analysis yields two functions: the autocorrelation function (ACF) and the partial autocorrelation function (PACF) \cite{brockwell2002introduction, brockwell2009time, parzen1963spectral}. These two metrics are depicted in Figure~\ref{fig:four_problem_trends_ACF_PACF} for the first (top left) time series of Figure~\ref{fig:four_problem_trends}, computed using the \emph{statsmodels} python package \cite{seabold2010statsmodels}. As seen in Figure~\ref{fig:four_problem_trends_ACF_PACF}, there is a clear autocorrelation and partial autocorrelation between the series, with up to around $40$ lags in the ACF plot (indicating an order of $40$ for the moving average), and up to around $10$ lags in the PACF plot (indicating an order of $10$ for the autoregressive model). This indicates that the observed time series exhibits a pattern that can be described by an appropriate ARMA model.

To test our hypothesis further, we also run an augmented Dickey-Fuller test \cite{DickeyFuller1979, mackinnon2010critical, mackinnon1994approximate, hamilton2020time} on the four time series datasets of Figure~\ref{fig:four_problem_trends} (without moving average applied) using the \emph{statsmodels} python package \cite{seabold2010statsmodels}. The Dickey–Fuller method tests the null hypothesis that a unit root is present in an autoregressive time series model, with the alternative being stationarity or trend-stationarity. When applying this test to the four datasets in Figure~\ref{fig:four_problem_trends}, we indeed reject the null in all four cases (p-values of 2.43e-25 and 2.97e-25 for the two MC experiments, respectively, and p-values of 2.68e-26 and 8.33e-26 for the two MVC problems, respectively), thus hinting at the existence of a trend in the datasets of Figure~\ref{fig:four_problem_trends}.

\subsection{Correlation between the energy time series of the problem QUBO and the performance indicator}
\label{sec:correlations}
Next, we establish that there is a significant correlation between the problem QUBO and the performance indicator. The following sections consider the three D-Wave models discussed in Section~\ref{sec:introduction} separately.

\subsubsection{DW\_2000Q\_LANL}
\label{sec:LANL}
\begin{figure}[h]
    \centering
    \includegraphics[align=c,width=0.8\textwidth]{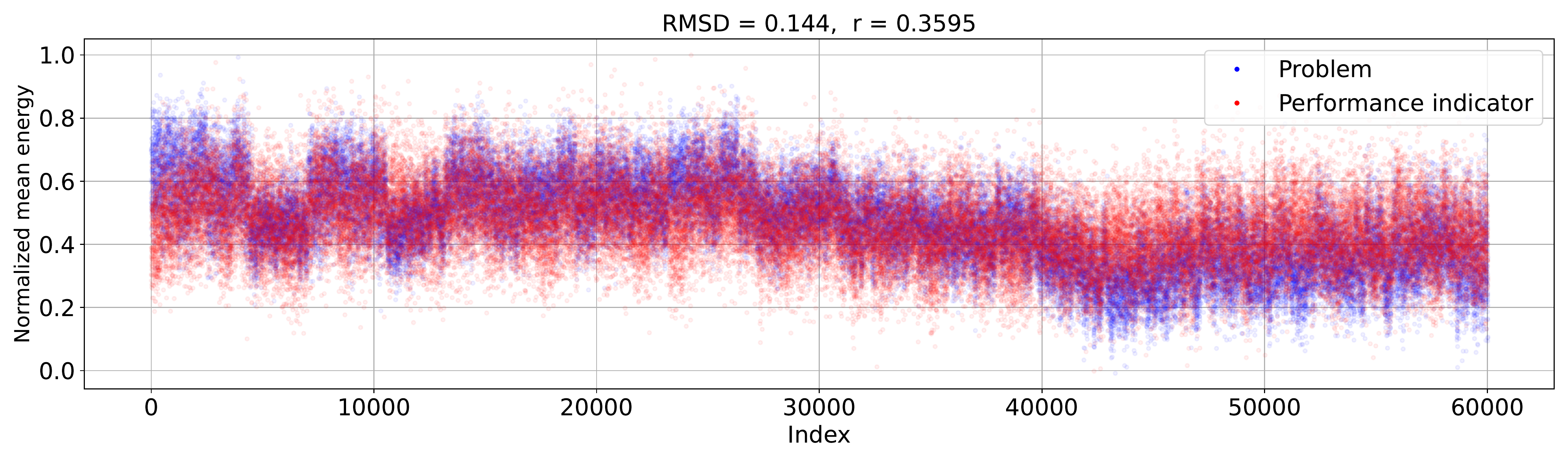}
    \includegraphics[align=c,width=0.19\textwidth]{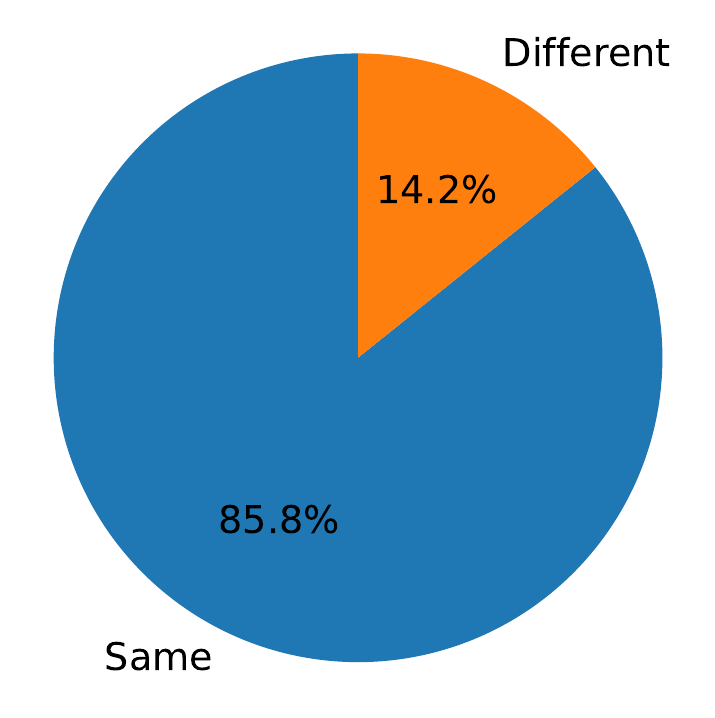}\\
    \includegraphics[align=c,width=0.8\textwidth]{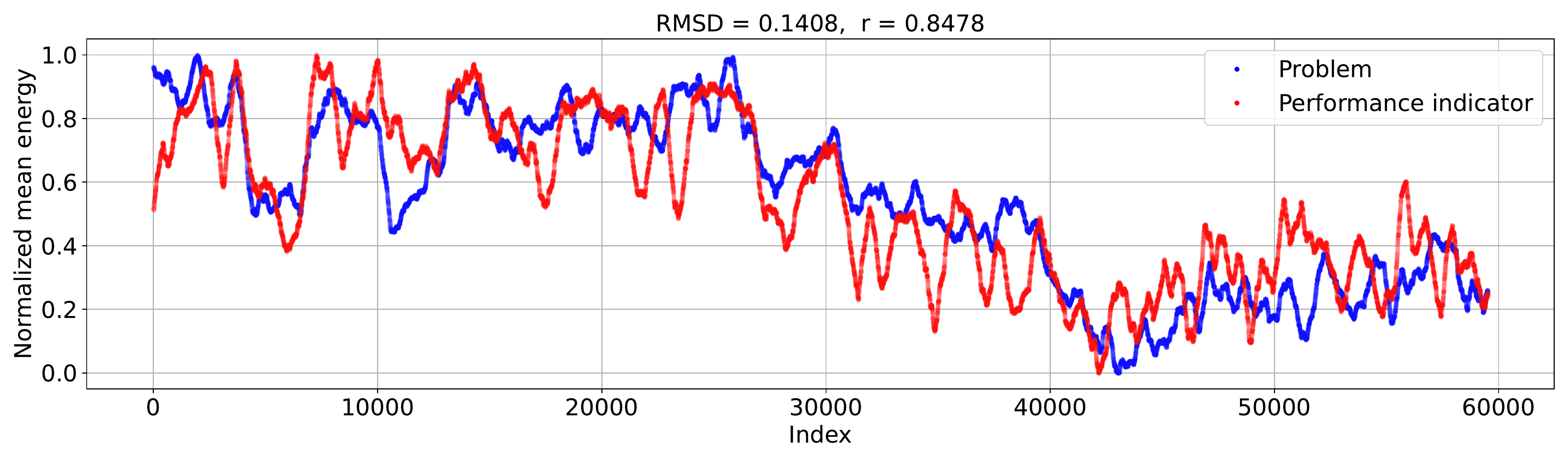}
    \includegraphics[align=c,width=0.19\textwidth]{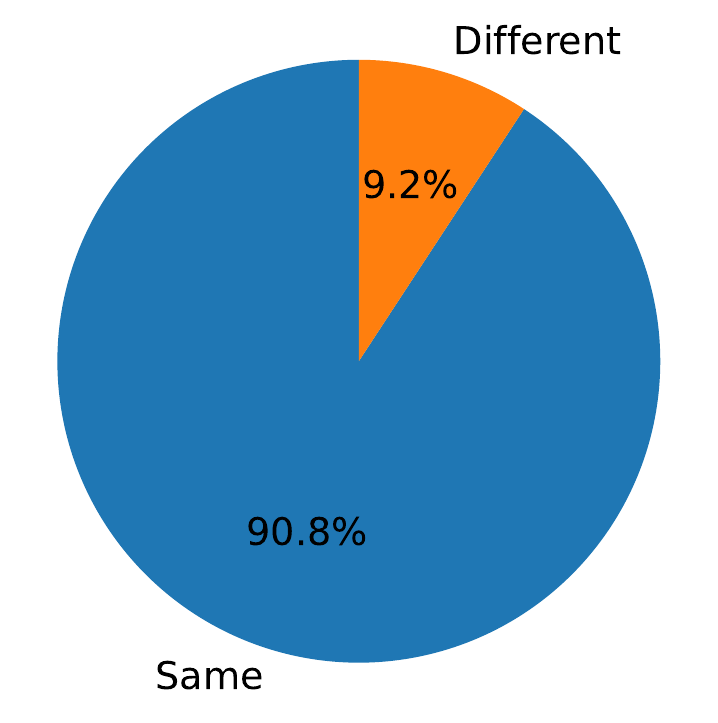}
    \caption{Observed mean (per backend job) normalized energies, represented as a time series, on \texttt{DW\_2000Q\_LANL} with an annealing time of $100$ microseconds for performance indicator PI1 without moving average (top) and with a moving average of window size $500$ (bottom). QUBO for the MC problem in blue and performance indicator in red. Pearson correlation and RMSD between the time-series for the problem QUBO and performance indicator are given on top of the subfigures. To the right of each of the time series plots are pie charts showing the percentage of the datapoints of the problem QUBO whose bin number matches the bin number predicted by the PI.}
    \label{fig:lanl1}
\end{figure}

We start by investigating the performance indicator PI1 (see Section~\ref{sec:indicator}) in connection with \texttt{DW\_2000Q\_LANL}. Figure~\ref{fig:lanl1}, left, shows the observed energies after unembedding for all $60,000$ D-Wave calls, without (top) and with a moving average (bottom).

Two observations are noteworthy. First, we observe that using a moving average is beneficial for highlighting the trend of the time series. Second, we see that indeed, after scaling the energy reads appropriately, the time series of the problem QUBO and the performance indicator roughly follow the same trend. This can be quantified through a high Pearson correlation coefficient and a low RMSD error between both series. This observation validates the proposed approach to use the performance indicator to monitor the sampling quality of the D-Wave annealer.

Apart from a visually comparing of the observed energies for both the problem QUBO and the PI, we also consider how well the performance indicator predicts the quality of the solution of the main problem. Specifically, we investigate whether the PI can distinguish between four quality levels: worst, bad, good, and best. To do this, we divide the normalized energies (the y-axis of Figure~\ref{fig:lanl1}) into four equally spaced bins (quartiles), representing the \textit{worst} solution quality ($y \in [0.75,1]$), \textit{bad} solution quality ($y \in [0.5,0.75)$), \textit{good} solution quality ($y \in [0.25,0.5)$), and \textit{best} solution quality ($y \in [0,0.25)$). For each data point, we can check whether the PI bin number correctly predicts the bin number of the problem QUBO. The pie charts in Figure~\ref{fig:lanl1} to the right of the time series graphs show the proportion of times that the $60,000$ datapoints of both the problem QUBO and the PI fell into the same bin (denoted by "Same") or different bins (denoted by "Different"). These pie charts provide an indication of how accurately the PI predicts the solution quality of the problem QUBO. The pie chart for the moving average plot (bottom plot in Figure~\ref{fig:lanl1}) is also computed using the moving averages data, and shows an increase in agreement between the PI and the problem QUBO compared to top pie chart.

\begin{figure}[ht]
    \centering
    \includegraphics[align=c,width=0.8\textwidth]{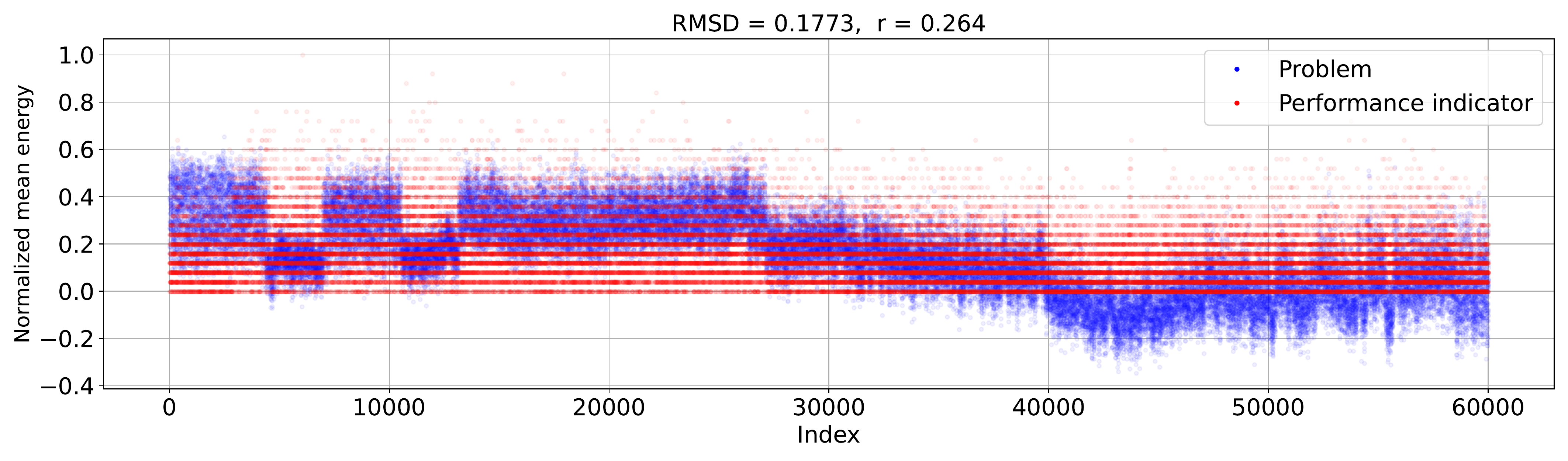}\hspace{0.2cm}
    \includegraphics[align=c,width=0.165\textwidth]{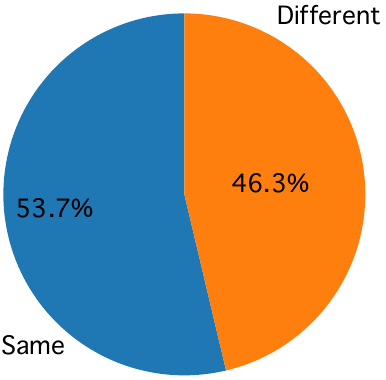}\\
    \includegraphics[align=c,width=0.8\textwidth]{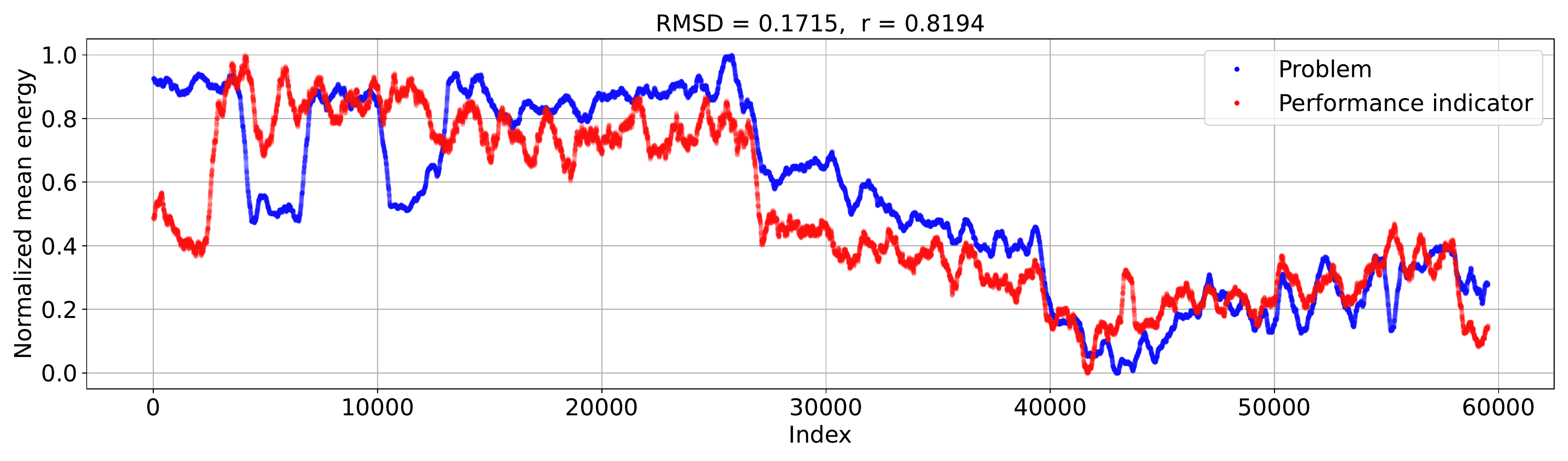}
    \includegraphics[align=c,width=0.19\textwidth]{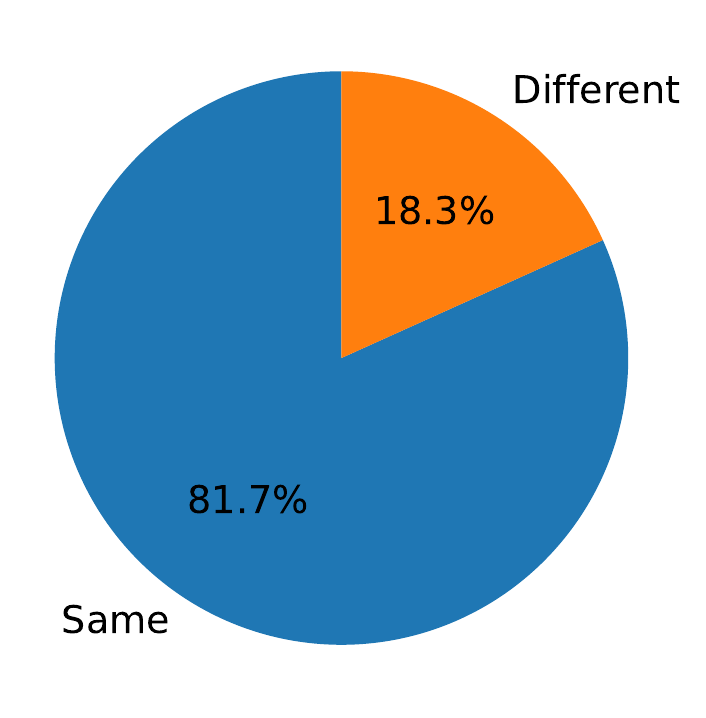}\\
    \caption{Observed mean normalized energies on \texttt{DW\_2000Q\_LANL} with an annealing time of $100$ microseconds for performance indicator PI2 without moving average (top) and with a moving average of window size $500$ (bottom). QUBO for the MC problem in blue and performance indicator in red. Pearson correlation and RMSD between the timeseries for the problem QUBO and performance indicator are given on top of the subfigures. To the right of each of the time series plots are pie charts showing the percentage of the datapoints of the problem QUBO whose bin number matches the bin number predicted by the PI.}
    \label{fig:lanl2}
\end{figure}

Figure~\ref{fig:lanl2} repeats the experiment using the performance indicator PI2, see Section~\ref{sec:indicator}, which only uses discrete QUBO weights $-1$ and $+1$. This is reflected in the reads for the problem QUBO (blue), which has a visually stratified distribution. However, a trend can be visualized again with the help of a moving average. As observed for Figure~\ref{fig:lanl1}, the problem QUBO and performance indicator roughly follow the same trend.

As the moving average allows us to better compare the time series observed for the problem QUBO and the performance indicator, we focus on the moving average figures in the remainder of this section.

\subsubsection{D-Wave 2000Q\_6}
\label{sec:2000Q}
\begin{figure}[ht]
    \centering
    \includegraphics[align=c,width=0.8\textwidth]{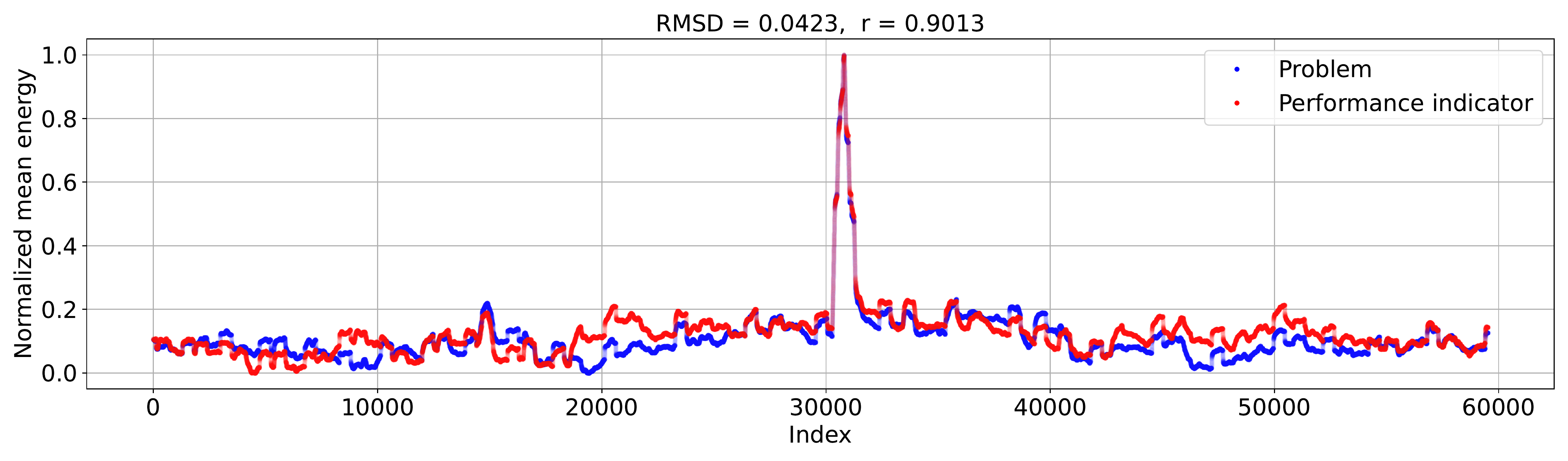}
    \includegraphics[align=c,width=0.19\textwidth]{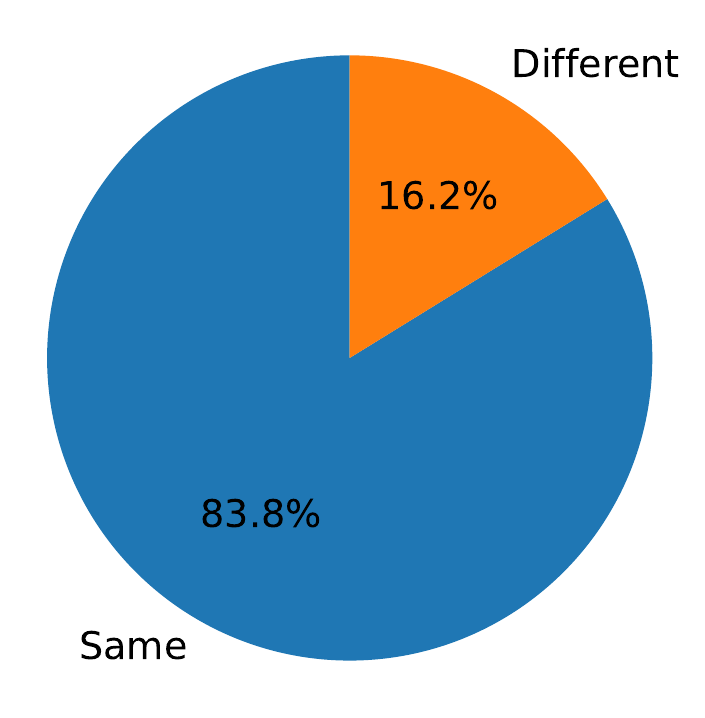}\\
    \includegraphics[align=c,width=0.8\textwidth]{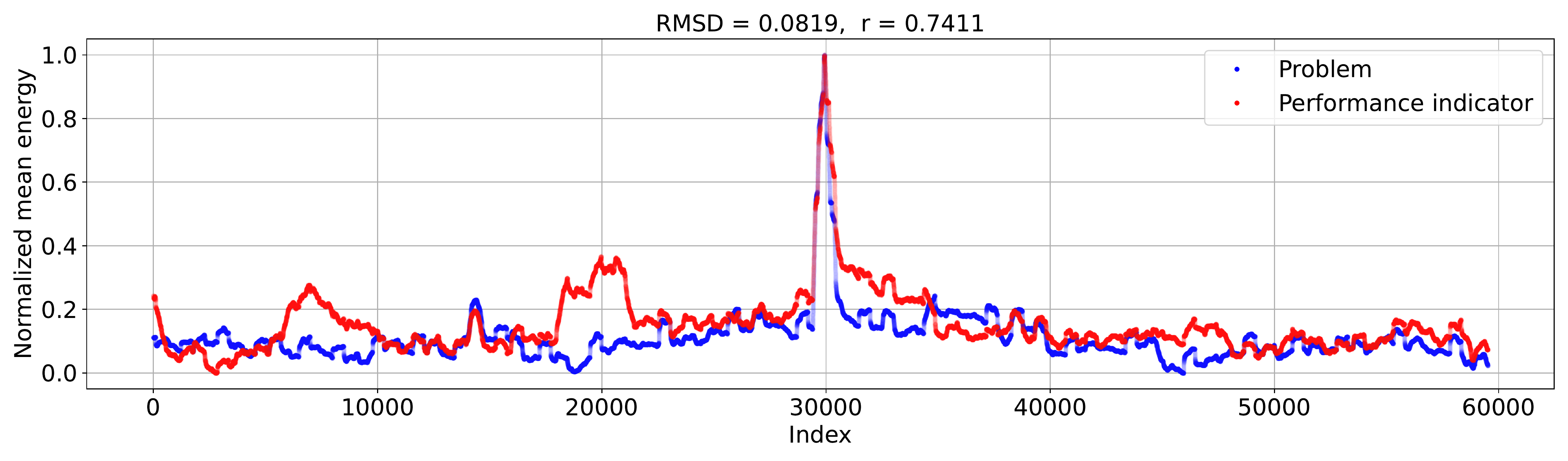}
    \includegraphics[align=c,width=0.19\textwidth]{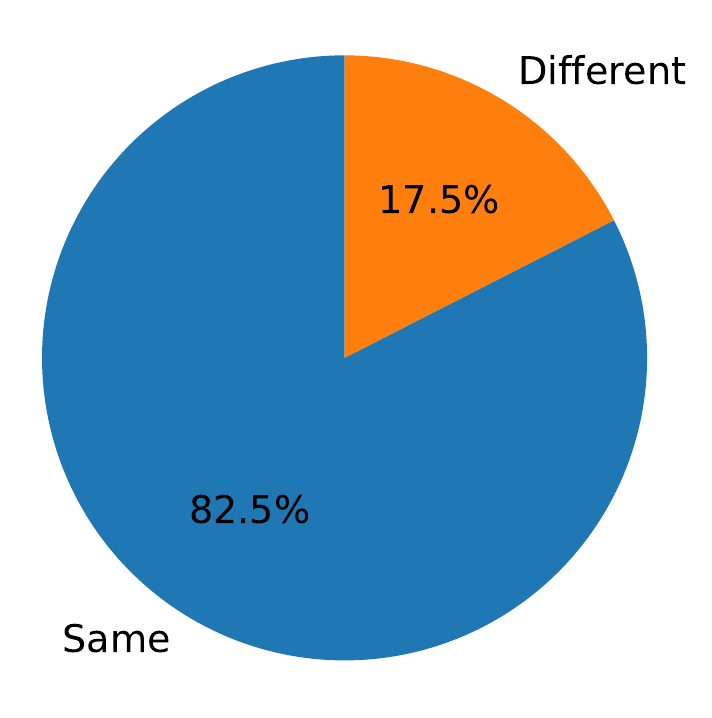}
    \caption{Observed normalized energies on \texttt{DW\_2000Q\_6} with an annealing time of $100$ microseconds for performance indicator PI1 (top) and PI2 (bottom) with a moving average of window size $500$. QUBO for the MC problem in blue and performance indicator in red. Pearson correlation and RMSD between the timeseries for the problem QUBO and performance indicator are given on top of the subfigures. To the right of each of the time series plots are pie charts showing the percentage of the datapoints of the problem QUBO whose bin number matches the bin number predicted by the PI.}
    \label{fig:2000Q}
\end{figure}

We repeat the same experiment on \texttt{DW\_2000Q\_6}, accessed through D-Wave Leap. Figure~\ref{fig:2000Q} shows the observed time series of energy reads after a post processing step with moving averages. The figure confirms that both performance indicators are a valid proxy for the behavior of the problem QUBO (the MC problem). Moreover, it appears that, in this experiment, the match is better for \texttt{DW\_2000Q\_6} than for \texttt{DW\_2000Q\_LANL}, and for performance indicator PI2 compared to PI1.

\subsubsection{Advantage\_system4.1}
\label{sec:advantage}
\begin{figure}[ht]
    \centering
    \includegraphics[align=c,width=0.8\textwidth]{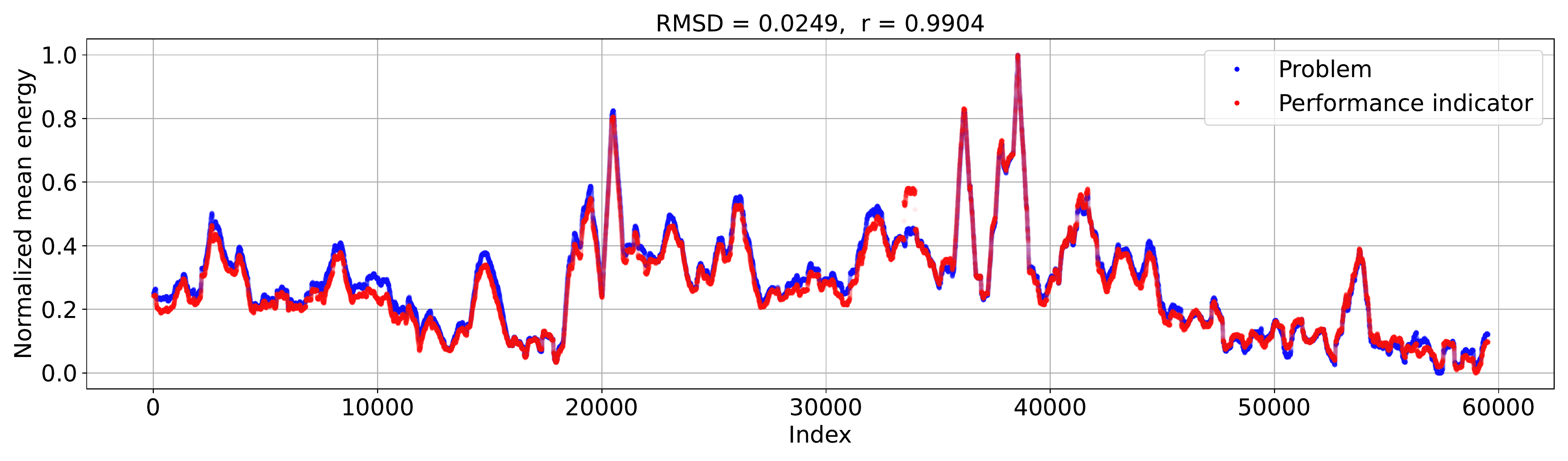}
    \includegraphics[align=c,width=0.19\textwidth]{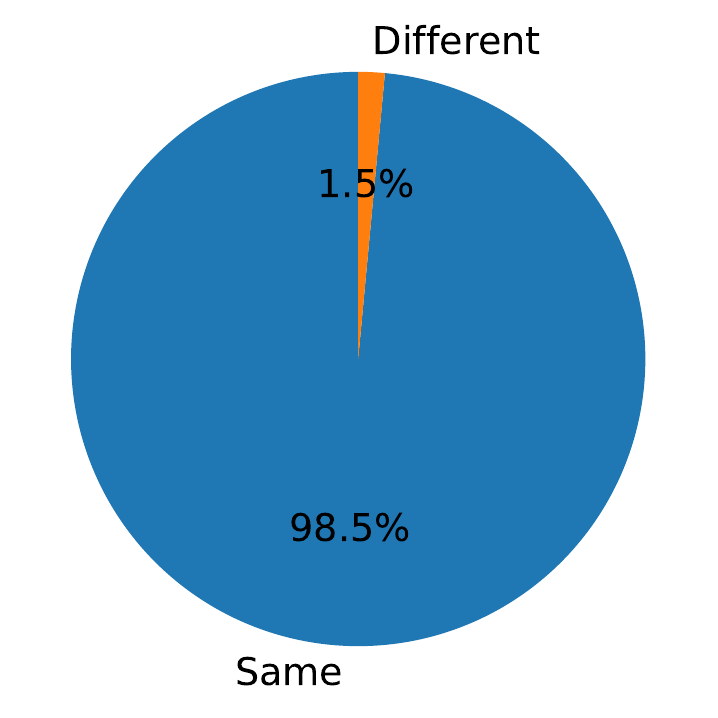}
    \includegraphics[align=c,width=0.8\textwidth]{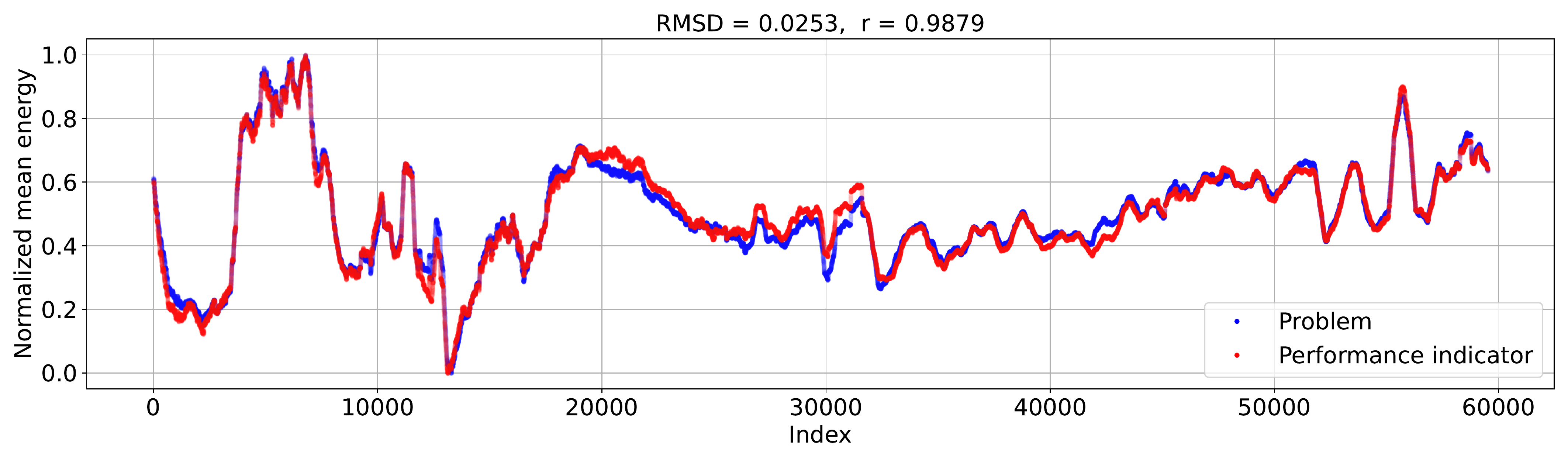}
    \includegraphics[align=c,width=0.19\textwidth]{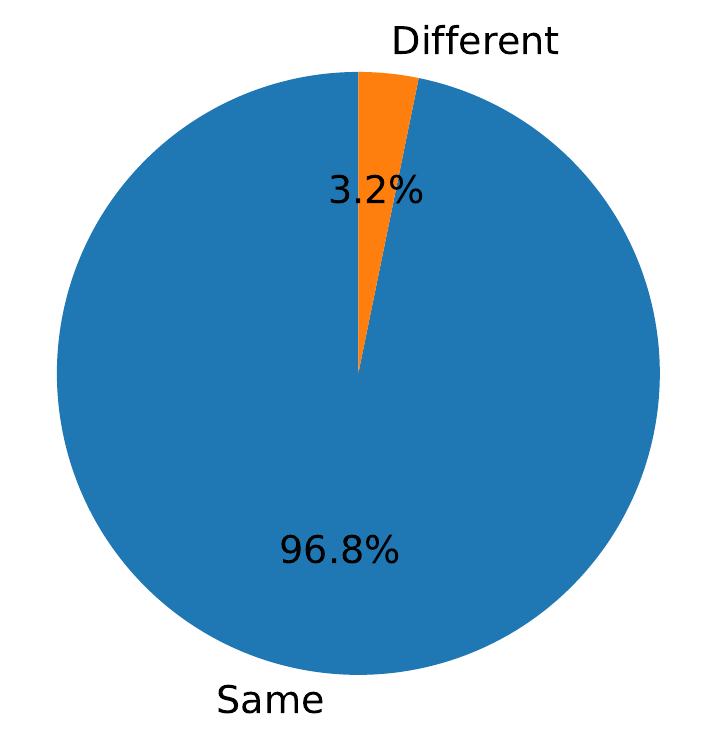}
    \caption{Observed normalized energies on \texttt{Advantage\_system4.1} with an annealing time of $100$ microseconds for performance indicator PI1 with a moving average of window size $500$ when the minor embedded MC problem has $80$ nodes (top) and $177$ nodes (bottom). QUBO for the MC problem in blue and performance indicator in red. Pearson correlation and RMSD between the timeseries for the problem QUBO and performance indicator are given on top of the subfigures. To the right of each of the time series plots are pie charts showing the percentage of the datapoints of the problem QUBO whose bin number matches the bin number predicted by the PI.}
    \label{fig:advantage}
\end{figure}

Finally, we again repeat the same experiment on Advantage\_system4.1 using performance indicator PI1 and two different instances of the MC problem of varying size. Figure~\ref{fig:advantage} shows the observed time series of energy reads after a moving average post processing step. The figure confirms that the performance indicator is a valid proxy for the behavior of the problem QUBO (the MC problem). Notably, on Advantage\_system4.1 the time series of the problem QUBO and performance indicator show a higher degree of similarity than on the previous D-Wave annealer systems (\texttt{DW\_2000Q\_LANL} and \texttt{DW\_2000Q\_6}).

\subsection{Stability of the performance indicator across problem instances}
\label{sec:alternating}
We aim to test whether the behavior of the performance indicator is stable across different instances of the MC problem. For this, we fix the performance indicator PI1 (see Section~\ref{sec:indicator}), and map the QUBOs of two different MC instances having the same number of nodes ($177$) onto the minor embedding in an alternating fashion, for a total of $5,000$ D-Wave calls. The alternating reads are chosen to ensure that we do not capture any time effects during the experiment.

\begin{figure}[ht]
    \centering
    \includegraphics[width=0.49\textwidth]{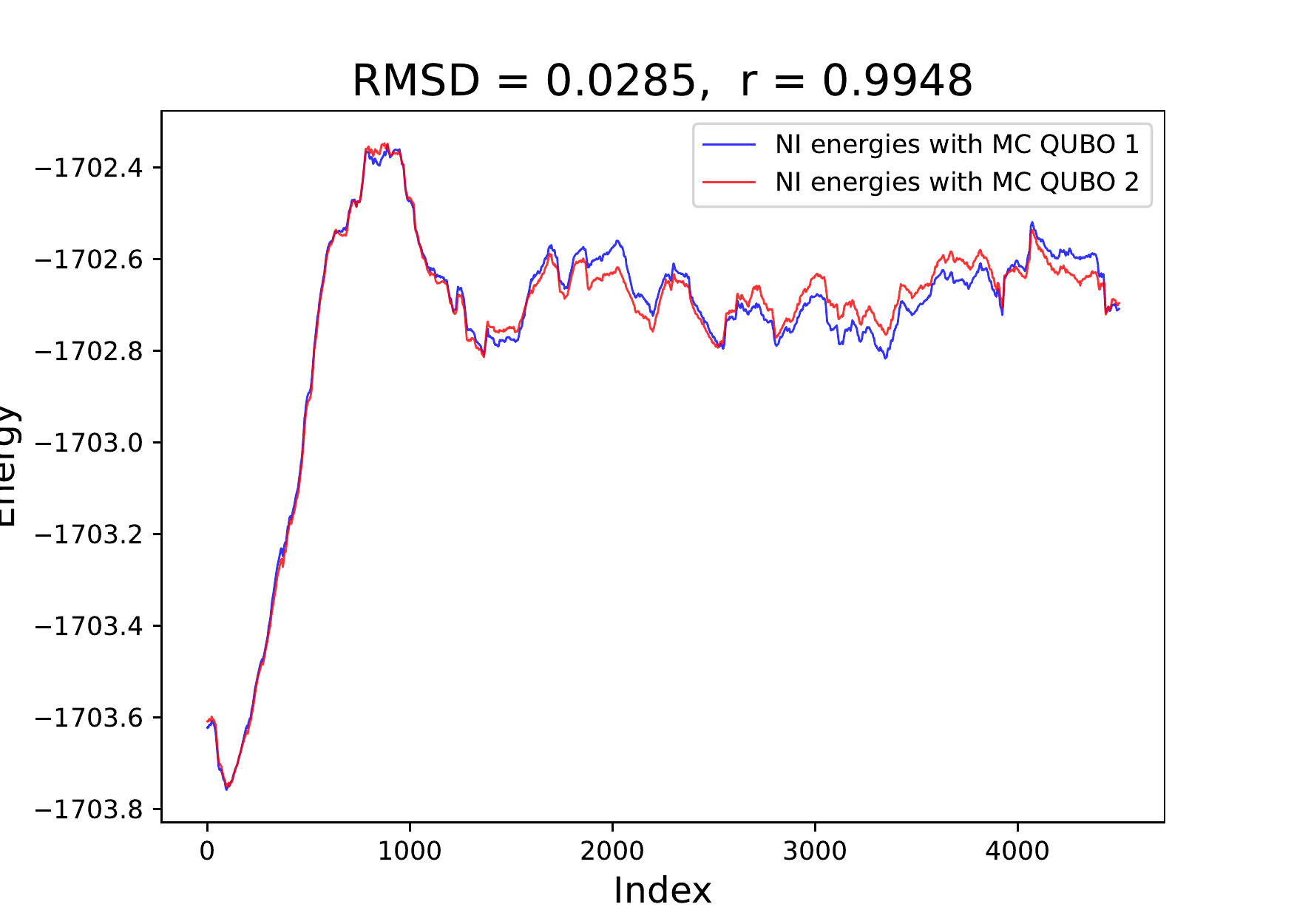}
    \includegraphics[width=0.49\textwidth]{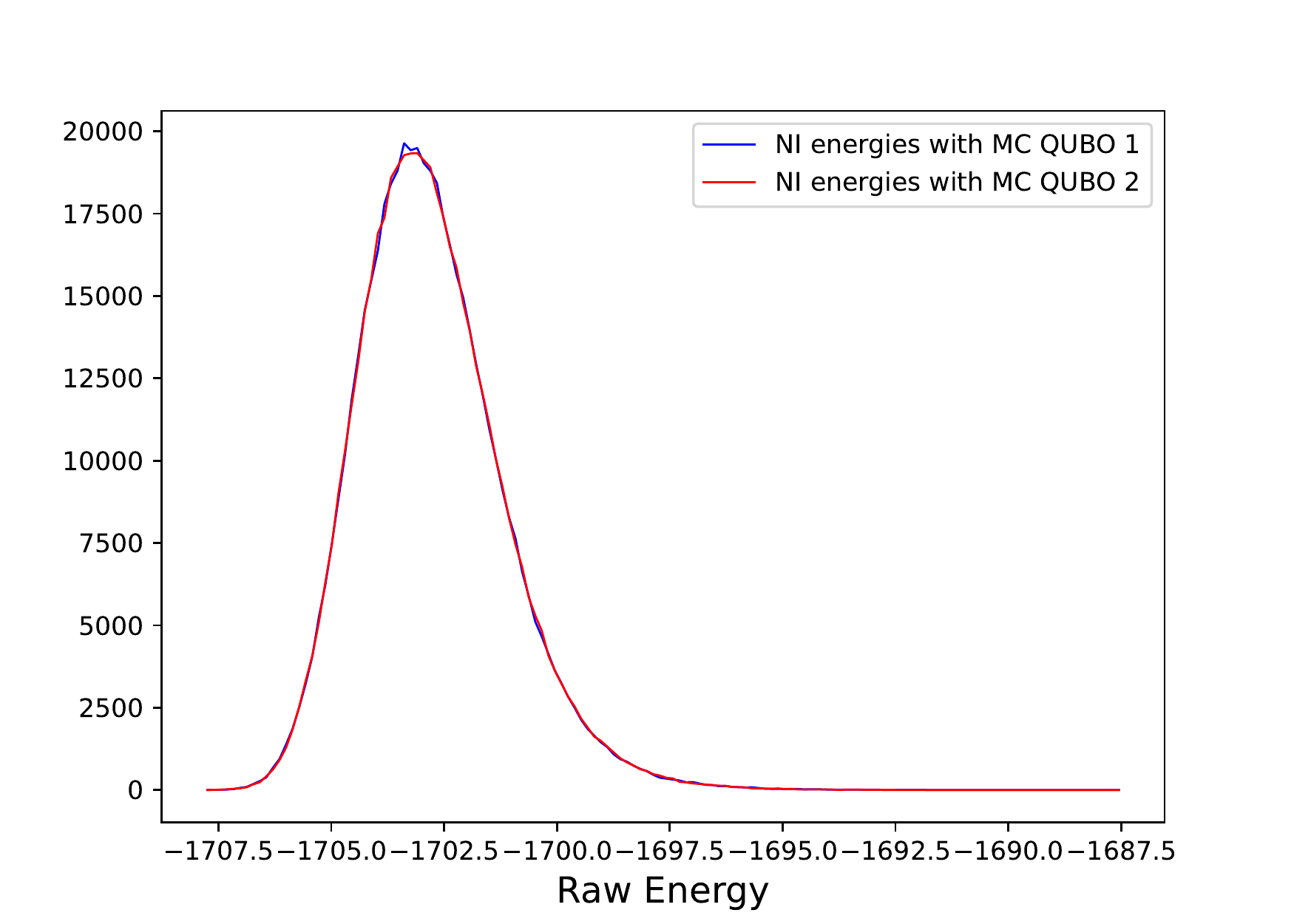}
    \caption{Solving two MC instances on random graphs of size $177$ while keeping the performance indicator PI1 fixed. Energy time series of $5,000$ D-Wave calls with moving average of window length $500$ (left), and distribution of the same energy reads (right). Results are from \texttt{Advantage\_system4.1}. }
    \label{fig:alternating_QUBOs}
\end{figure}

We read out the observed energies for the performance indicator, and record if each energy read was observed in connection with the first or the second MC QUBO. We visualize both the time series (with moving average of window size $500$) as well as the distribution of the observed energies in Figure~\ref{fig:alternating_QUBOs}. As can be seen in the figure, the distribution of reads is indeed visually indistinguishable. A two sample Kolmogorov-Smirnov test conducted with the function \textit{kstest} in Python Scipy \cite{scipy, Hodges1958} reveals a p-value of roughly $0.54$, indicating a failure to reject the null hypothesis of the distributions being different. The result shows that indeed, using a fixed performance indicator is a valid approach to monitor the fluctuations of the D-Wave annealer for various problem QUBOs being solved.

\subsection{Using the performance indicator to assess the quality of a solution}
\label{sec:histograms}
Finally, we aim to use the performance indicator to help us quantify the quality of the samples obtained from the D-Wave quantum annealer. A simple algorithm to attempt this was presented in Section~\ref{sec:algo}. In this section, we verify that the reads for the problem QUBO are indeed of better quality when sampling at times when the performance indicator suggests that D-Wave is in a phase of lower noise.

\begin{figure}[ht]
    \centering
    \includegraphics[width=0.8\textwidth]{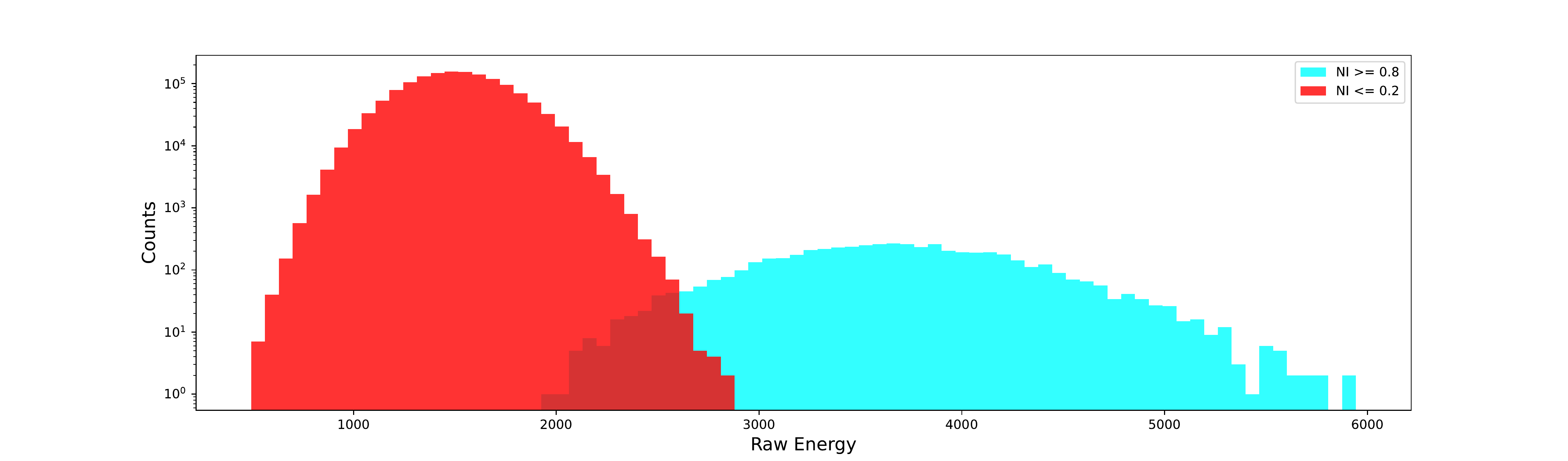}
    \caption{Histogram of the observed energies for the MC QUBO at time points when the normalized energy value $e$ (normalized to lay within $[0,1]$) obtained from the performance indicator is $e \leq 0.2$ (red) or $e \geq 0.8$ (blue), see Section~\ref{sec:algo}. Experiment uses performance indicator PI1. Results are from \texttt{Advantage\_system4.1}. }
    \label{fig:histogram_energy_difference}
\end{figure}

Figure~\ref{fig:histogram_energy_difference} shows a histogram of samples obtained for the problem QUBO, stratified into two sets: those that are obtained at times when the normalized energy $e \in [0,1]$ of the performance indicator, see Section~\ref{sec:algo}, satisfies $e \leq 0.2$ and once when it satisfies $e \geq 0.8$. Figure~\ref{fig:histogram_energy_difference} is computed cumulatively, using the procedure described in Section~\ref{sec:algo}, with a burn-in period $b=10$, and therefore provides a straightforward example of how the performance indicator works. We observe that indeed, the samples for the problem QUBO read out at times when the performance indicator suggested a low-noise state have considerably lower energy values (meaning they are of better quality and, on average, closer to the ground state of eq.~(\ref{eq:hamiltonian}).

\section{Discussion}
\label{sec:discussion}
In this paper we demonstrate that the performance of the most recent D-Wave quantum annealers exhibits long-term trends. Moreover, we demonstrate that unused hardware qubits can be used to measure the current relative performance state and to improve the quality of the solutions to an optimization problem.

This was accomplished by fixing a complete (clique) embedding to implement a given problem QUBO (in this article we considered the MC problem) onto the D-Wave quantum hardware, and by utilizing the remaining hardware qubits to implement another QUBO with random weights (which we call the performance indicator). Not only do we observe a distinct correlation between the performance indicator and the embedded problem QUBO, but we also observe a considerable trends in the solution quality over time across all three quantum annealers. 

We observed that the newest D-Wave generation, \texttt{Advantage\_system4.1}, yields a higher degree of similarity between the time series for the problem QUBO and the performance indicator. A possible reason explaining why there is a poorer correlation on the previous devices \texttt{DW\_2000Q\_LANL} and \texttt{DW\_2000Q\_6} compared to the newest Advantage\_system4.1 could lay in the fact that the \texttt{DW\_Advantage} hardware offers a much denser connectivity than the previous generations, which allows for a better connectivity of the unused qubits left over for the performance indicator.

This work leaves scope for a variety of future research avenues:
\begin{enumerate}
    \item One could take advantage of the time series correlations in the quantum annealing noise profiles in order to train machine learning models to predict the future noise profile of the performance indicator. This could be a way to allow users to determine when the device is performing relatively well without needing to execute extra jobs on the backend. 
    \item One could apply change point detection methods \cite{polunchenko2012state, aminikhanghahi2017survey} to the time series observed for the performance indicator in order to determine longer periods of time within which the D-Wave quantum annealer is performing sub-optimally, thus allowing the user to wait some period of time before continuing to execute jobs.
    \item Unlike circuit model quantum computers, current quantum annealers do not have succinct error rate characterizations available to users. However, there has been a recent algorithm proposed (QASA) which provides single qubit fidelity assessment for quantum annealers \cite{Nelson2021}. Quantifying how the single qubit fidelity characterization drifts over time (i.e., the trends observed in this paper) across all qubits would be an interesting research avenue. 
    \item Consider a performance indicator composed entirely of zero coefficient Isings on the remaining native hardware graph. In an ideal quantum annealing computation, this would not yield interesting trends. However, because of spin bath polarization and environmental noise, even reading out the idle qubit states for programmed coefficients of $0$ on the unused parts of the hardware could provide a valid performance indicator metric.
    \item Perform a more comprehensive study on how different problem types and annealing parameters, such as annealing time, influence the two-phase thresholding algorithm performance indicator accuracy. 
\end{enumerate}

\section*{Acknowledgments}
\label{sec:acknowledgments}
This work was supported by the U.S. Department of Energy through the Los Alamos National Laboratory. Los Alamos National Laboratory is operated by Triad National Security, LLC, for the National Nuclear Security Administration of U.S. Department of Energy (Contract No.\ 89233218CNA000001). The research presented in this article was supported by the Laboratory Directed Research and Development program of Los Alamos National Laboratory under project number 20220656ER. The work of Hristo Djidjev has been also partially supported by Grant No.\ BG05M2OP001-1.001-0003, financed by the Science and Education for Smart Growth Operational Program (2014-2020) and co-financed by the European Union through the European structural and Investment funds and by Grant No KP-06-DB-11 of the Bulgarian National Science Fund. This research used resources provided by the Darwin testbed at Los Alamos National Laboratory (LANL) which is funded by the Computational Systems and Software Environments subprogram of LANL's Advanced Simulation and Computing program (NNSA/DOE). The authors thank Aaron Barbosa for his help with the analysis of the time series data.

\noindent
LA-UR-22-27371

\appendix
\section{Summary table of conducted experiments}
\label{sec:appendix}
All experiments run in the scope of this work are summarized in Table~\ref{table:all_results}. In Table~\ref{table:all_results}, experiments 1, 2, and 3 show that the Pearson correlation on \texttt{DW\_2000Q\_LANL} decreases as annealing time increases, and that RMSD decreases as annealing time increases. This observation is confirmed by experiments 4, 5, and 6. On \texttt{Advantage\_system4.1}, the same observation also holds true for the Pearson correlation while the RMSD stays roughly unchanged as the annealing time increases (see experiments $9-12$ and $15-17$).

\begin{table}
\begin{center}
\begin{tabular}{|l|l|l|l|l|l|l|}
    \hline
    No. & PI & $R^2$ & RMSD & Chip ID & Anneal time & Embedding size\\
    \hline
    1 & 1 & 0.4512 & 0.1423 & DW\_2000Q\_LANL & AT 1 & 65\\ 
    2 & 1 & 0.3595 & 0.144 & DW\_2000Q\_LANL & AT 100 & 65\\ 
    3 & 1 & 0.2928 & 0.0629 & DW\_2000Q\_LANL & AT 1000 & 65\\ 
    \hline
    4 & 2 & 0.2983 & 0.1798 & DW\_2000Q\_LANL & AT 1 & 65\\ 
    5 & 2 & 0.264 & 0.1773 & DW\_2000Q\_LANL & AT 100 & 65\\ 
    6 & 2 & 0.1875 & 0.1628 & DW\_2000Q\_LANL & AT 1000 & 65\\ 
    \hline
    7 & 1 & 0.9171 & 0.0316 & DW\_2000Q\_LANL & random AT & 65\\ 
    8 & 2 & 0.1009 & 0.1218 & DW\_2000Q\_LANL & random AT & 65\\ 
    \hline
    9 & 1 & 0.9726 & 0.0336 & Advantage\_system4.1 & AT 20 & 80\\
    10 & 2 & 0.9458 & 0.0474 & Advantage\_system4.1 & AT 20 & 80\\ 
    11 & 1 & 0.802 & 0.0745 & Advantage\_system4.1 & AT 100 & 80\\
    12 & 2 & 0.8237 & 0.0866 & Advantage\_system4.1 & AT 100 & 80\\
    \hline
    13 & 1 & 0.8032 & 0.0437 & DW\_2000Q\_6 & AT 100 & 50\\
    14 & 2 & 0.5762 & 0.0645 & DW\_2000Q\_6 & AT 100 & 50\\
    \hline
    15 & 1 & 0.9602 & 0.0199 & Advantage\_system4.1 & AT 100 & 177\\
    16 & 2 & 0.9404 & 0.0293 & Advantage\_system4.1 & AT 100 & 177\\ 
    17 & 1 & 0.9143 & 0.0292 & Advantage\_system4.1 & AT 100 & 80\\ 
    \hline
\end{tabular}
\end{center}
\caption{Summary of all experiments being conducted in this work. Number of experiment (column~1), performance indicator QUBO being used (column~2), Pearson correlation (column~3), root-mean-square deviation (RMSD) between the problem QUBO and the performance indicator QUBO (column~4), Chip ID of D-Wave device (column~5), annealing time (AT) (column~6), and size of the clique embedding (column~7). All numbers are rounded to four decimal places.}
\label{table:all_results}
\end{table}

\clearpage

\setlength\bibitemsep{0pt}
\printbibliography

@article{Preskill2018quantumcomputingin,
    doi = {10.22331/q-2018-08-06-79},
    url = {https://doi.org/10.22331/q-2018-08-06-79},
    title = {Quantum {C}omputing in the {NISQ} era and beyond},
    author = {Preskill, J.},
    journal = {{Quantum}},
    issn = {2521-327X},
    publisher = {{Verein zur F{\"{o}}rderung des Open Access Publizierens in den Quantenwissenschaften}},
    volume = {2},
    pages = {79},
    month = aug,
    year = {2018}
}

@unpublished{Boothby2020,
    author = {Boothby, K. and Bunyk, P. and Raymond, J. and Roy, A.},
    title = {{Next-Generation Topology of D-Wave Quantum Processors}}, 
    year = {2020},
    note = {arXiv:2003.00133}
}

@misc{https://doi.org/10.48550/arxiv.1901.07636,
    doi = {10.48550/ARXIV.1901.07636},  
    url = {https://arxiv.org/abs/1901.07636},
    author = {Dattani, N. and Szalay, S. and Chancellor, N.},
    title = {Pegasus: The second connectivity graph for large-scale quantum annealing hardware},
    publisher = {arXiv},
    year = {2019}
}

@unpublished{Dasgupta2021,
    title = {Stability of noisy quantum computing devices},
    author = {Dasgupta, S. and Humble, T.},
    year = {2021},
    note = {arXiv:2105.09472}
}

@article{Zaborniak2021,
    author = {Zaborniak, T. and de Sousa, R.},
    title = {{Benchmarking Hamiltonian Noise in the D-Wave Quantum Annealer}},
    year = {2021},
    journal = {IEEE Transactions on Quantum Engineering},
    volume = {2},
    pages = {1–6},
    doi = {10.1109/tqe.2021.3050449}
}

@article{Nelson2021,
    title = {Single-qubit fidelity assessment of quantum annealing hardware},
    author = {Nelson, J. and Vuffray, M. and Lokhov, A. and Coffrin, C.},
    journal = {IEEE Transactions on Quantum Engineering},
    year = {2021},
    volume = {2},
    pages = {1--10},
    doi={10.1109/TQE.2021.3092710}
}

@inproceedings{Zaborniak2020,
    author = {Zaborniak, T. and de Sousa, R.},
    booktitle = {2020 IEEE International Conference on Quantum Computing and Engineering (QCE)},
    title = {{In Situ Noise Characterization of the D-Wave Quantum Annealer}},
    year = {2020},
    pages = {409-412},
    doi={10.1109/QCE49297.2020.00057}
}

@article{Pelofske2022parallel,
    title = {Parallel quantum annealing}, 
    author = {Pelofske, E. and Hahn, G. and Djidjev, H.},
    year = {2022},
    journal = {Sci Rep},
    volume = {12},
    pages = {4499},
	doi = {10.1038/s41598-022-08394-8}
}

@unpublished{Cai2014,
    title = {A practical heuristic for finding graph minors}, 
    author = {Cai, J. and Macready, W. and Roy, A.},
    year = {2014},
    note = {arXiv:1406.2741}
}

@article{Barahona1982,
    title = {On the computational complexity of Ising spin glass models},
    author = {Barahona, F.},
    journal = {Journal of Physics A: Mathematical and General},
    volume = {15},
    number = {10},
    pages = {3241},
    year = {1982}
}

@article{Lucas2014,
    title = {{Ising formulations of many NP problems}},
    author = {Lucas, A.},
    journal = {Front Physics},
    volume = {2},
    number = {5},
    year = {2014}
}

@inproceedings{Chapuis2017,
    author = {Chapuis, G. and Djidjev, H. and Hahn, G. and Rizk, G.},
    title = {{Finding Maximum Cliques on a Quantum Annealer}},
    year = {2017},
    booktitle = {Proceedings of the Computing Frontiers Conference},
    pages = {63-70}
}

@article{Pelofske2019mc,
    title = {{Solving large Maximum Clique problems on a quantum annealer}}, 
    author = {Pelofske, E. and Hahn, G. and Djidjev, H.},
    journal = {{Proceedings of the First International Workshop on Quantum Technology and Optimization Problems QTOP'19}},
    year = {2019}
}

@unpublished{Yarkoni2022,
    author = {Yarkoni, S. and Raponi, E. and Schmitt, S. and B{\"a}ck, T.},
    title = {{Quantum Annealing for Industry Applications: Introduction and Review}},
    note = {arXiv:2112.07491},
    year = {2022}
}

@misc{scipy,
    author = {Gommers, R. and Virtanen, P. and Burovski, E. and Weckesser, W. and Oliphant, T. and Haberland, T. and Cournapeau, D. and Reddy, T. and alexbrc and Peterson, P. and Nelson, A. and Wilson, A. and endolith and Mayorov, N. and Polat, I. and van der Walt, S. and Laxalde, D. and Brett, M. and Roy, P. and Larson, E. and Millman, J. and Lars and peterbell10 and van Mulbregt, P. and Sakai, A. and Carey, C. and eric-jones and Kern, R. and Kai},
    title = {SciPy v1.8.1},
    year = {2022},
    month = {May},
    url = {https://doi.org/10.5281/zenodo.6560517}
}

@article{Hodges1958,
    author = {Hodges, J.},
    title = {{The significance probability of the Smirnov two-sample test}},
    journal = {Arkiv f{\"o}r Matematik},
    volume = {3},
    number = {5},
    pages = {469-486},
    year = {1958}
}

@misc{matplotlib,
    author = {Caswell, T. and Droettboom, M. and Lee, A. and Sales de Andrade, E. and Hoffmann, T. and Klymak, J. and Hunter, J. and Firing, E. and Stansby, D. and Varoquaux, N. and Nielsen, J. and Root, B. and May, R. and Elson, P. and Sepp{\"a}nen, J. and Dale, D. and Lee, J. and McDougall, D. and Straw, A. and Hobson, P. and hannah and Gohlke, C. and Vincent, A. and Yu, T. and Ma, E. and Silvester, S. and Moad, C. and Kniazev, N. and Ernest, E. and Ivanov, P.},
    title = {Matplotlib v3.5.2},
    year = {2022},
    month = {May},
    url = {https://doi.org/10.5281/zenodo.6513224}
}

@article{Hunter2007,
    author = {Hunter, J.},
    title = {{Matplotlib: A 2D graphics environment}},
    journal = {Computing in Science \& Engineering},
    volume = {9},
    number = {3},
    pages = {90--95},
    year = {2007}
}

@inproceedings{networkx,
    author = {Hagberg, A. and Schult, D. and Swart, P.},
    title = {{Exploring network structure, dynamics, and function using NetworkX}},
    booktitle = {Proceedings of the 7th Python in Science Conference (SciPy2008), G{\"a}el Varoquaux, Travis Vaught, and Jarrod Millman (Eds.), Pasadena, CA USA},
    pages = {11-15},
    year = {2008}
}

@article{Raymond2016,
    author = {Raymond, J. and Yarkoni, S. and Andriyash, E.},
    title = {{Global Warming: Temperature Estimation in Annealers}},
    journal = {Front ICT},
    year = {2016},
    volume = {3},
    number = {23}
}

@unpublished{Andriyash2017,
    author = {Andriyash, E. and Amin, M.},
    title = {{Can quantum Monte Carlo simulate quantum annealing?}},
    year = {2017},
    note = {arXiv:1703.09277}
}

@article{Denchev2016,
    author = {Denchev, V. and Boixo, S. and Isakov, S. and Ding, N. and Babbush, R. and Smelyanskiy, V. and Martinis, J. and Neven, H.},
    title = {{What is the Computational Value of Finite-Range Tunneling?}},
    journal = {Phys Rev X},
    volume = {6},
    number = {031015},
    year = {2016}
}

@unpublished{ocean,
    author = {{D-Wave Systems}},
    title = {{Ocean SDK}},
    year = {2022},
    note = {\url{https://docs.ocean.dwavesys.com/en/stable/index.html}}
}

@article{Harper2020,
    title = {Efficient learning of quantum noise},
    author = {Harper, R. and Flammia, S. and Wallman, J.},
    year = {2020},
    journal = {Nature Physics},
    volume = {16},
    pages = {1184-1188}
}

@inproceedings{Hamilton2020,
    author = {Hamilton, K. and Kharazi, T. and Morris, T. and McCaskey, A. and Bennink, R. and Pooser, R.},
    title = {Scalable quantum processor noise characterization},
    booktitle = {2020 IEEE International Conference on Quantum Computing and Engineering (QCE)},
    year = {2020},
    pages = {430-440}
}

@unpublished{Suzuki2020,
    author = {Suzuki, T. and Nakazato, H.},
    title = {A proposal of noise suppression for quantum annealing},
    note = {arXiv:2006.13440},
    year = {2020}
}

@unpublished{Shaib2021,
    author = {Shaib, A. and Naim, M. and Fouda, M. and Kanj, R. and Kurdahi, F.},
    title = {{Efficient Noise Mitigation Technique for Quantum Computing}},
    note = {arXiv:2109.05136},
    year = {2021}
}

@article{Vinci2016,
    author = {Vinci, W. and Albash, T. and Lidar, D.},
    title = {Nested quantum annealing correction},
    journal = {npj Quantum Information},
    year = {2016},
    volume = {2},
    number = {16017}
}

@article{Pearson2019,
    author = {Pearson, A. and Mishra, A. and Hen, I. and Lidar, D.},
    title = {Analog errors in quantum annealing: doom and hope},
    journal = {npj Quantum Information},
    year = {2019},
    volume = {5},
    number = {107}
}

@article{Devitt2013,
    author = {Devitt, S. and Munro, W. and Nemoto, K.},
    title = {Quantum error correction for beginners},
    year = {2013},
    journal = {Rep. Prog. Phys.},
    volume = {76},
    number = {076001}
}

@article{Pudenz2014,
    author = {Pudenz, K. and Albash, T. and Lidar, D.},
    title = {Error-corrected quantum annealing with hundreds of qubits},
    journal = {Nature Communications},
    volume = {5},
    number = {3243},
    year = {2014}
}

@article{PhysRevA.91.042302,
    title = {Quantum annealing correction for random Ising problems},
    author = {Pudenz, K.L. and Albash, T. and Lidar, D.A.},
    journal = {Phys. Rev. A},
    volume = {91},
    issue = {4},
    pages = {042302},
    numpages = {15},
    year = {2015},
    month = {Apr},
    publisher = {American Physical Society},
    doi = {10.1103/PhysRevA.91.042302},
    url = {https://link.aps.org/doi/10.1103/PhysRevA.91.042302}
}

@article{Roffe2019,
    author = {Roffe, J.},
    title = {Quantum error correction: an introductory guide},
    journal = {Contemporary Physics},
    volume = {60},
    number = {3},
    year = {2019}
}

@misc{1503.05679,
    author = {Perdomo-Ortiz, A. and O'Gorman, B. and Fluegemann, J. and Biswas, R. and Smelyanskiy, V.N.},
    title = {{D}etermination and correction of persistent biases in quantum annealers},
    year = {2015},
    eprint = {1503.05679},
    note = {arXiv:1503.05679v1}
}

@misc{https://doi.org/10.48550/arxiv.1507.04774,
    doi = {10.48550/ARXIV.1507.04774},  
    url = {https://arxiv.org/abs/1507.04774},
    author = {Boothby, K. and King, A.D. and Roy, A.},
    title = {Fast clique minor generation in Chimera qubit connectivity graphs},
    publisher = {arXiv},
    year = {2015}
}

@article{morita2008mathematical,
    title={Mathematical foundation of quantum annealing},
    author={Morita, S. and Nishimori, H.},
    journal={Journal of Mathematical Physics},
    volume={49},
    number={12},
    pages={125210},
    year={2008},
    publisher={American Institute of Physics}
}

@article{RevModPhys.80.1061,
    title = {Colloquium: Quantum annealing and analog quantum computation},
    author = {Das, A. and Chakrabarti, B.K.},
    journal = {Rev. Mod. Phys.},
    volume = {80},
    issue = {3},
    pages = {1061--1081},
    numpages = {0},
    year = {2008},
    month = {Sep},
    publisher = {American Physical Society},
    doi = {10.1103/RevModPhys.80.1061},
    url = {https://link.aps.org/doi/10.1103/RevModPhys.80.1061}
}

@article{hauke2020perspectives,
    title={Perspectives of quantum annealing: Methods and implementations},
    author={Hauke, P. and Katzgraber, H.G. and Lechner, W. and Nishimori, H. and Oliver, W.D.},
    journal={Reports on Progress in Physics},
    volume={83},
    number={5},
    pages={054401},
    year={2020},
    publisher={IOP Publishing},
	doi = {10.1088/1361-6633/ab85b8}
}

@article{finnila1994quantum,
    title={Quantum annealing: A new method for minimizing multidimensional functions},
    author={Finnila, A.B. and Gomez, M.A. and Sebenik, C. and Stenson, C. and Doll, J.D.},
    journal={Chemical physics letters},
    volume={219},
    number={5-6},
    pages={343--348},
    year={1994},
    publisher={Elsevier}
}

@article{kadowaki1998quantum,
    title={Quantum annealing in the transverse Ising model},
    author={Kadowaki, T. and Nishimori, H.},
    journal={Physical Review E},
    volume={58},
    number={5},
    pages={5355},
    year={1998},
    publisher={APS}
}

@article{johnson2011quantum,
    title={Quantum annealing with manufactured spins},
    author={Johnson, M.W. and Amin, M. and Gildert, S. and Lanting, T. and Hamze, F. and Dickson, N. and Harris, R. and Berkley, A.J. and Johansson, J. and Bunyk, P. and others},
    journal={Nature},
    volume={473},
    number={7346},
    pages={194--198},
    year={2011},
    publisher={Nature Publishing Group}
}

@article{pelofske2022quantum,
    title={Quantum annealing algorithms for Boolean tensor networks},
    author={Pelofske, E. and Hahn, G. and O’Malley, D. and Djidjev, H.N. and Alexandrov, B.S.},
    journal={Scientific Reports},
    volume={12},
    number={1},
    pages={1--19},
    year={2022},
    publisher={Nature Publishing Group},
    doi = {10.1038/s41598-022-12611-9}
}

@article{PhysRevLett.116.220501,
    title = {Mean Field Analysis of Quantum Annealing Correction},
    author = {Matsuura, S. and Nishimori, H. and Albash, T. and Lidar, D.A.},
    journal = {Phys. Rev. Lett.},
    volume = {116},
    issue = {22},
    pages = {220501},
    numpages = {5},
    year = {2016},
    month = {Jun},
    publisher = {American Physical Society},
    doi = {10.1103/PhysRevLett.116.220501},
    url = {https://link.aps.org/doi/10.1103/PhysRevLett.116.220501}
}

@article{vinci2015quantum,
    title={Quantum annealing correction with minor embedding},
    author={Vinci, W. and Albash, T. and Paz-Silva, G. and Hen, I. and Lidar, D.A.},
    journal={Physical Review A},
    volume={92},
    number={4},
    pages={042310},
    year={2015},
    publisher={APS}
}

@misc{https://doi.org/10.48550/arxiv.2108.11334,
    doi = {10.48550/ARXIV.2108.11334},  
    url = {https://arxiv.org/abs/2108.11334},
    author = {Suau, A. and Nelson, J. and Vuffray, M. and Lokhov, A.Y. and Cincio, L. and Coffrin, C.},
    title = {Single-Qubit Cross Platform Comparison of Quantum Computing Hardware},
    publisher = {arXiv},
    year = {2021}
}

@article{alexander2020qiskit,
    title={Qiskit pulse: Programming quantum computers through the cloud with pulses},
    author={Alexander, T. and Kanazawa, N. and Egger, D.J. and Capelluto, L. and Wood, C.J. and Javadi-Abhari, A. and McKay, D.C.},
    journal={Quantum Science and Technology},
    volume={5},
    number={4},
    pages={044006},
    year={2020},
    publisher={IOP Publishing},
	doi = {10.1088/2058-9565/aba404}
}

@inproceedings{9283531,
    author={Wood, C.J.},
    booktitle={2020 IEEE 38th International Conference on Computer Design (ICCD)}, 
    title={Special Session: Noise Characterization and Error Mitigation in Near-Term Quantum Computers}, 
    year={2020},
    pages={13-16},
    doi={10.1109/ICCD50377.2020.00016}
}

@article{sanders2015bounding,
    title={Bounding quantum gate error rate based on reported average fidelity},
    author={Sanders, Y.R. and Wallman, J.J. and Sanders, B.C.},
    journal={New Journal of Physics},
    volume={18},
    number={1},
    pages={012002},
    year={2015},
    publisher={IOP Publishing}
}

@article{aminikhanghahi2017survey,
    title={A survey of methods for time series change point detection},
    author={Aminikhanghahi, S. and Cook, D.J.},
    journal={Knowledge and information systems},
    volume={51},
    number={2},
    pages={339--367},
    year={2017},
    publisher={Springer}
}

@article{polunchenko2012state,
    title={State-of-the-art in sequential change-point detection},
    author={Polunchenko, A.S. and Tartakovsky, A.G.},
    journal={Methodology and computing in applied probability},
    volume={14},
    number={3},
    pages={649--684},
    year={2012},
    publisher={Springer}
}

@inproceedings{Zbinden2020,
    title = {Embedding Algorithms for Quantum Annealers with Chimera and Pegasus Connection Topologies},
    author = {Zbinden, S. and B{\"a}rtschi, A. and Djidjev, H. and Eidenbenz, S.},
    booktitle = {Sadayappan, P., Chamberlain, B., Juckeland, G., Ltaief, H. (eds) High Performance Computing. ISC High Performance 2020. Lecture Notes in Computer Science},
    volume = {12151},
    published = {Springer, Cham.},
    year = {2020},
    doi={10.1007/978-3-030-50743-5_10}
}

@article{ayanzadeh2021multi,
    title={Multi-qubit correction for quantum annealers},
    author={Ayanzadeh, R. and Dorband, J. and Halem, M. and Finin, T.},
    journal={Scientific Reports},
    volume={11},
    number={1},
    pages={1--12},
    year={2021},
    publisher={Nature Publishing Group},
    doi={10.1038/s41598-021-95482-w}
}

@unpublished{spin-bath-polarization,
    author = {{D-Wave Systems}},
    title = {{Spin-Bath Polarization Effect}},
    year = {2022},
    note = {\url{https://docs.dwavesys.com/docs/latest/c_qpu_errors.html}}
}

@article{PhysRevApplied.15.014029,
    title = {Comparing Relaxation Mechanisms in Quantum and Classical Transverse-Field Annealing},
    author = {Albash, T. and Marshall, J.},
    journal = {Phys. Rev. Applied},
    volume = {15},
    issue = {1},
    pages = {014029},
    numpages = {14},
    year = {2021},
    month = {Jan},
    publisher = {American Physical Society},
    doi = {10.1103/PhysRevApplied.15.014029},
    url = {https://link.aps.org/doi/10.1103/PhysRevApplied.15.014029}
}

@misc{uniform-torque-compensation,
    title = {uniform torque compensation},
    year = {2022},
    author = {D-Wave Systems Inc.}, 
    url = {https://docs.ocean.dwavesys.com/projects/system/en/stable/reference/generated/dwave.embedding.chain_strength.uniform_torque_compensation.html}
}

@article{lobe2021minor,
    title={{Minor Embedding in Broken Chimera and Pegasus Graphs is NP-complete}},
    author={Lobe, E. and Lutz, A.},
    journal={arXiv preprint arXiv:2110.08325},
    year={2021}
}

@article{PhysRevA.98.022314,
    title = {Reverse annealing for the fully connected $p$-spin model},
    author = {Ohkuwa, M. and Nishimori, H. and Lidar, D.A.},
    journal = {Phys. Rev. A},
    volume = {98},
    issue = {2},
    pages = {022314},
    numpages = {12},
    year = {2018},
    month = {Aug},
    publisher = {American Physical Society},
    doi = {10.1103/PhysRevA.98.022314},
    url = {https://link.aps.org/doi/10.1103/PhysRevA.98.022314}
}

@article{DickeyFuller1979,
    author = {Dickey, D.A. und Fuller, W.A.},
    title = {{Distribution of the Estimators for Autoregressive Time Series with a Unit Root}},
    journal = {Journal of the American Statistical Association},
    year = {1979},
    volume = {74},
    pages = {427-431}
}

@inproceedings{seabold2010statsmodels,
  title={statsmodels: Econometric and statistical modeling with python},
  author={Seabold, Skipper and Perktold, Josef},
  booktitle={9th Python in Science Conference},
  year={2010},
}

@techreport{mackinnon2010critical,
  title={Critical values for cointegration tests},
  author={MacKinnon, James G},
  year={2010},
  institution={Queen's Economics Department Working Paper}
}

@article{mackinnon1994approximate,
  title={Approximate asymptotic distribution functions for unit-root and cointegration tests},
  author={MacKinnon, James G},
  journal={Journal of Business \& Economic Statistics},
  volume={12},
  number={2},
  pages={167--176},
  year={1994},
  publisher={Taylor \& Francis}
}

@book{hamilton2020time,
  title={Time series analysis},
  author={Hamilton, James Douglas},
  year={2020},
  publisher={Princeton university press}
}

@book{brockwell2002introduction,
  title={Introduction to time series and forecasting},
  author={Brockwell, Peter J and Davis, Richard A},
  year={2002},
  publisher={Springer}
}

@book{brockwell2009time,
  title={Time series: theory and methods},
  author={Brockwell, Peter J and Davis, Richard A},
  year={2009},
  publisher={Springer science \& business media}
}

@article{parzen1963spectral,
  title={On spectral analysis with missing observations and amplitude modulation},
  author={Parzen, Emanuel},
  journal={Sankhy{\=a}: The Indian Journal of Statistics, Series A},
  pages={383--392},
  year={1963},
  publisher={JSTOR}
}

@misc{https://doi.org/10.48550/arxiv.2209.03796,
  doi = {10.48550/ARXIV.2209.03796},
  
  url = {https://arxiv.org/abs/2209.03796},
  
  author = {Mineh, Lana and Montanaro, Ashley},
  
  title = {Accelerating the variational quantum eigensolver using parallelism},
  
  publisher = {arXiv},
  
  year = {2022},
  
  copyright = {arXiv.org perpetual, non-exclusive license}
}

@misc{https://doi.org/10.48550/arxiv.2102.05321,
  doi = {10.48550/ARXIV.2102.05321},
  
  url = {https://arxiv.org/abs/2102.05321},
  
  author = {Niu, Siyuan and Todri-Sanial, Aida},
  
  title = {Enabling multi-programming mechanism for quantum computing in the NISQ era},
  
  publisher = {arXiv},
  
  year = {2021},
  
  copyright = {arXiv.org perpetual, non-exclusive license}
}

@INPROCEEDINGS{9407180,

  author={Liu, Lei and Dou, Xinglei},

  booktitle={2021 IEEE International Symposium on High-Performance Computer Architecture (HPCA)}, 

  title={QuCloud: A New Qubit Mapping Mechanism for Multi-programming Quantum Computing in Cloud Environment}, 

  year={2021},

  volume={},

  number={},

  pages={167-178},

  doi={10.1109/HPCA51647.2021.00024}}

@inproceedings{10.1145/3352460.3358287,
author = {Das, Poulami and Tannu, Swamit S. and Nair, Prashant J. and Qureshi, Moinuddin},
title = {A Case for Multi-Programming Quantum Computers},
year = {2019},
isbn = {9781450369381},
publisher = {Association for Computing Machinery},
address = {New York, NY, USA},
url = {https://doi.org/10.1145/3352460.3358287},
doi = {10.1145/3352460.3358287},
booktitle = {Proceedings of the 52nd Annual IEEE/ACM International Symposium on Microarchitecture},
pages = {291–303},
numpages = {13},
location = {Columbus, OH, USA},
series = {MICRO '52}
}

@misc{https://doi.org/10.48550/arxiv.2112.00387,
  doi = {10.48550/ARXIV.2112.00387},
  
  url = {https://arxiv.org/abs/2112.00387},
  
  author = {Niu, Siyuan and Todri-Sanial, Aida},
  
  title = {How Parallel Circuit Execution Can Be Useful for NISQ Computing?},
  
  publisher = {arXiv},
  
  year = {2021},
  
  copyright = {arXiv.org perpetual, non-exclusive license}
}

@misc{https://doi.org/10.48550/arxiv.2205.12165,
  doi = {10.48550/ARXIV.2205.12165},
  
  url = {https://arxiv.org/abs/2205.12165},
  
  author = {Pelofske, Elijah and Hahn, Georg and Djidjev, Hristo N.},
  
  title = {Solving Larger Optimization Problems Using Parallel Quantum Annealing},
  
  publisher = {arXiv},
  
  year = {2022},
  
  copyright = {arXiv.org perpetual, non-exclusive license}
}

@article{proctor2020detecting,
  title={Detecting and tracking drift in quantum information processors},
  author={Proctor, Timothy and Revelle, Melissa and Nielsen, Erik and Rudinger, Kenneth and Lobser, Daniel and Maunz, Peter and Blume-Kohout, Robin and Young, Kevin},
  journal={Nature communications},
  volume={11},
  number={1},
  pages={1--9},
  year={2020},
  publisher={Nature Publishing Group}
}

@ARTICLE{9749894,

  author={Ohkura, Yasuhiro and Satoh, Takahiko and Van Meter, Rodney},

  journal={IEEE Transactions on Quantum Engineering}, 

  title={Simultaneous Execution of Quantum Circuits on Current and Near-Future NISQ Systems}, 

  year={2022},

  volume={3},

  number={},

  pages={1-10},

  doi={10.1109/TQE.2022.3164716}}

@article{ahsan2020reconfiguring,
  title={Reconfiguring quantum error-correcting codes for real-life errors},
  author={Ahsan, Muhammad and Naqvi, Syed Abbas Zilqurnain},
  journal={Journal of Physics D: Applied Physics},
  volume={53},
  number={41},
  pages={415302},
  year={2020},
  publisher={IOP Publishing},
  doi={10.1088/1361-6463/ab96eb}
}

@article{Ahsan_2022,
	doi = {10.1103/physreva.105.022428},
  
	url = {https://doi.org/10.1103%2Fphysreva.105.022428},
  
	year = 2022,
	month = {feb},
  
	publisher = {American Physical Society ({APS})},
  
	volume = {105},
  
	number = {2},
  
	author = {Muhammad Ahsan and Syed Abbas Zilqurnain Naqvi and Haider Anwer},
  
	title = {Quantum circuit engineering for correcting coherent noise},
  
	journal = {Physical Review A}
}

@article{PhysRevX.4.021041,
  title = {Entanglement in a Quantum Annealing Processor},
  author = {Lanting, T. and Przybysz, A. J. and Smirnov, A. Yu. and Spedalieri, F. M. and Amin, M. H. and Berkley, A. J. and Harris, R. and Altomare, F. and Boixo, S. and Bunyk, P. and Dickson, N. and Enderud, C. and Hilton, J. P. and Hoskinson, E. and Johnson, M. W. and Ladizinsky, E. and Ladizinsky, N. and Neufeld, R. and Oh, T. and Perminov, I. and Rich, C. and Thom, M. C. and Tolkacheva, E. and Uchaikin, S. and Wilson, A. B. and Rose, G.},
  journal = {Phys. Rev. X},
  volume = {4},
  issue = {2},
  pages = {021041},
  numpages = {14},
  year = {2014},
  month = {May},
  publisher = {American Physical Society},
  doi = {10.1103/PhysRevX.4.021041},
  url = {https://link.aps.org/doi/10.1103/PhysRevX.4.021041}
}

\end{document}